\documentclass[11pt]{article}

\usepackage{abstract}
\usepackage[utf8]{inputenc}
\usepackage[T1]{fontenc}
\usepackage[british]{babel}
\usepackage{graphicx}
\usepackage{amsmath, amssymb, bm}
\usepackage{booktabs}
\usepackage[breaklinks]{hyperref}
\usepackage[font={small,it}]{caption}	
\usepackage{subcaption}
\usepackage[export]{adjustbox}
\usepackage{booktabs}
\usepackage{makecell}
\usepackage{natbib}
\usepackage{abstract}
\usepackage{indentfirst}
\usepackage{comment}
\usepackage{xcolor}
\usepackage{algorithm}
\usepackage{algpseudocode}
\usepackage{algcompatible}
\usepackage{hyphenat}

\graphicspath{{img-horizon/}}

\usepackage[a4paper, margin=1in]{geometry}

\begin{document}
%
%\title{Attribute-driven Density Estimation for Density-based Community Detection}
\title{A density-based framework for community detection in attributed networks}

\author{Sara Geremia\textsuperscript{1}, Michael Fop\textsuperscript{2}, Domenico De Stefano\textsuperscript{1} \\[1ex]
\small \textsuperscript{1}Department of Political and Social Sciences, University of Trieste, Italy \\
\small \textsuperscript{2}School of Mathematics and Statistics, University College Dublin, Ireland
}
\date{}

\maketitle              % typeset the title of the contribution

\begin{abstract}

Community structure in social and collaborative networks often emerges from a complex interplay between structural mechanisms, such as degree heterogeneity and leader-driven attraction, and homophily on node attributes. %This interplay is particularly evident in large-scale research collaboration networks, such as those arising from European funding programmes, which are sparse, highly heterogeneous, and shaped by thematic priorities and leading organisations. 
Existing community detection methods typically focus on these dimensions in isolation, limiting their ability to recover interpretable communities in presence of such mechanisms.
In this paper, we propose AttDeCoDe, an attribute-driven extension of a density-based community detection framework, developed to analyse networks where node characteristics play a central role in group formation. Instead of defining density purely from network topology, AttDeCoDe estimates node-wise density in the attribute space, allowing communities to form around attribute-based community representatives while preserving structural connectivity constraints. This approach naturally captures homophily-driven aggregation while remaining sensitive to leader influence.
We evaluate the proposed method through a simulation study based on a novel generative model that extends the degree-corrected stochastic block model by incorporating attribute-driven leader attraction, reflecting key features of collaborative research networks. %Results show that AttDeCoDe achieves competitive performance relative to topology-based and attribute-assisted benchmarks.
We perform an empirical application to research collaboration data from the Horizon programmes, where organisations are characterised by project-level thematic descriptors. %, further demonstrates that the method identifies communities that are both structurally cohesive and thematically interpretable, revealing leader-centred cores embedded within attribute-coherent groups.
Both results show that AttDeCoDe offers a flexible and interpretable framework for community detection in attributed networks achieving competitive performance relative to topology-based and attribute-assisted benchmarks.
%, well suited to large-scale collaboration systems shaped by the joint influence of leadership and homophily.

\noindent \textbf{Keywords:} community detection, attributed networks, leader influence, homophily
\end{abstract}

\clearpage

\section{Introduction}

In social communities, individuals tend to form close relationships with other subjects who share similar traits or attract their interests, resulting in the formation of groups. These groups are typically characterised by denser or more intense connections among their members than between different groups. This structural pattern is commonly referred to as community structure \citep{Newman2004FindingNetworks} and can emerge through different social attachment mechanisms, possibly occurring concurrently or as alternative processes. Triadic closure is the most studied of these mechanisms and represents a natural tendency of real social networks to create group connections \citep{Bianconi2014TriadicCA}, involving trust as a potential aggregation driver. Other mechanisms can give rise to community structure by shaping how new ties are formed, leading to the emergence of cohesive subgroups

One mechanism is the attraction exerted by influential actors or network leaders, a process determined by the heterogeneity of degree among individuals \citep{Perc2014TheData}. Highly connected individuals may attract others, promoting the growth of subgroups centred around them.
Within these subgroups, leader nodes assume critical roles in the diffusion of information, ideas, and innovations. %These actors are essential for maintaining communication flow and network cohesion. 
Consequently, identifying the most influential members within communities has become a key research focus in network science. 
%Traditional algorithms typically detect communities first and then identify the leader or central nodes within them \cite{Rashidi2024}. 
%More recent leader-based approaches invert this process by first detecting leader nodes based on their structural influence and subsequently forming communities around them \citep{Yakoubi2014LICOD:Networks,Helal2017LeaderbasedNetworks,Ahajjam2018ANetworks,Lu2021Leader-BasedNetworks,Akachar2025LeaDCD:Networks}. This leader-first methodology captures the nature of community formation in many social systems, where influence often precedes group consolidation.

Another, complementary (or even alternative) aggregation mechanism is the propensity for individuals with similar interests or characteristics to bond, known as homophily. Homophily on observed attributes plays a relevant role in shaping the structure and dynamics of networks, contributing to the formation of communities \citep{Mcpherson2001BIRDSNetworks}. %Moreover, node attributes--such as demographic details or user-generated content--often provide valuable information about homopholic preferences; however, traditional community detection approaches have frequently overlooked this additional information \citep{Chunaev2020CommunitySurvey}.

%To address this limitation, several recent contributions have proposed integrating structural and attribute information to improve community detection. Notably, \cite{Chunaev2020CommunitySurvey} provides a useful taxonomy of how structure and attributes can be fused. The approaches collectively demonstrate that incorporating attribute information can significantly refine the identification of community boundaries, especially when one data source (structure or attributes) is incomplete or noisy.

The mentioned mechanisms do not exist in isolation. Rather, group formation in real social networks could emerge from a complex interplay between structural mechanisms, including triadic closure, preferential attachment associated with degree heterogeneity, and homophily on node attributes \citep{bachmann2025network}. Few contributions propose unified and interpretable frameworks that integrate them jointly—particularly in empirical settings characterised by sparsity, degree heterogeneity, and the availability of rich node attributes. Most studies, focusing on the definition of communities as dense groups, mainly explore the interaction between triadic closure and homophily \citep{triadichomophily,Peixoto2022DisentanglingNetworks, Mosleh2025}. However, growing empirical and theoretical evidence suggests that degree heterogeneity—through the emergence of hubs and bridging actors—plays a central role in shaping communities, interacting with homophily in ways that can either reinforce or blur community boundaries \citep{bachmann2025network,FUHSE202443,KOZITSIN2023100248}.
%Instead of an explicit bias towards popularity, preferential attachment can also be an implicit consequence of triadic closure, since well-connected individuals are more likely to appear as friends-of-friends \citep{bachmann2025network}.
Recent studies on leader-based community detection have demonstrated that, when considering the similarity of attributes together with the presence of leaders, the results tend to be more satisfactory, mainly because the attractiveness of leader nodes may depend on the characteristics of neighbouring nodes \citep{Helal2016AnNetworks,Lu2021Leader-BasedNetworks,Hu2024Network-adjustedDetection}. 

Building on the literature on leader-based community detection, and moving beyond a definition of communities based solely on density and triadic closure, we argue that leader-driven attraction and homophily jointly shape a structure in which multiple dense, leader-centred cores coexist.
%Such an architecture may emerge in degree-heterogeneous networks, in which nodes tend to cluster by shared attributes yet remain interconnected through bridging actors. 
Examples of networks exhibiting this pattern include citation networks, where disciplinary proximity generates subgroups led by prominent scholars. For instance, \cite{NewmanClusteringPA} noted that scientific collaboration and citation networks are governed by star scientists, while also exhibiting high levels of triadic closure. %in our opinion, homophily--in terms of research topics--can represent a possible driver. 

More generally, scientific collaboration networks provide a key empirical setting for investigating complex group formation mechanisms within the so-called science of science field \citep{liu2023data,fortunato2018science}. In particular, collaboration networks based on funded research projects exhibit structures that suggest the joint effect of leader-driven attraction and homophily: collaborative ties tend to form around thematic priorities and are often organised by leading organisations that act as coordination hubs \citep{Morea2024MappingAnalysis}. The identification of such key players--companies, academic institutions, or research organisations--is crucial for other organisations seeking to engage with them in future projects or for policy makers to evaluate the effectiveness of the funded research in a given region or around a specific topic \citep{Morea2024MappingAnalysis}.

Motivated by an original dataset on collaborative research, this work extends the analysis of community formation in leader-influenced networks to settings where node-level attributes are available.
In our framework, leaders are defined as nodes located in high-density regions of the attribute space, that is, nodes whose thematic profiles are representative of subpopulations. This perspective allows homophily to be captured naturally: the leaders' high attribute density identifies focal points around which other nodes with comparable characteristics cluster.

 Following a density-based community detection (DeCoDe) approach \citep{Menardi2022Density-basedNetworks}, we operationalise this idea by searching for dense subgraphs that form around attribute-defined leaders. This integration enables the detection of cohesive, attribute-consistent communities that reflect structural connectivity and homophily on observed attributes.

The rest of the paper is structured as follows.
In Section~\ref{sec:background}, we provide the empirical and theoretical background and review the most relevant literature.
Section~\ref{sec:method} presents the network structure under investigation, formally states the problem, and illustrates the proposed methodology.
Section~\ref{sec:simulation} describes the setup and outcomes of the simulation experiments, while Section~\ref{sec:application} describes the motivating network and reports the results of the application.
Finally, Section~\ref{sec:conclusion} offers concluding remarks and suggests directions for future work.

\section{Relevant literature}\label{sec:background}

\subsection{Empirical background}

During the last decades, complex networks have been recognised to be crucial in explaining social phenomena using the structural and relational features of the network of actors involved. %In many fields, the importance of the relational aspect of social phenomena and of social capital has been highlighted. 
In the innovation economics and related public interventions, knowledge and innovation networks are currently of great interest.

Social network analysis has been recognised as a fundamental tool in studying scientific collaboration in the science of science field \citep{liu2023data,fortunato2018science} and inter\hyp{}organisational interaction in terms of knowledge and innovation flows \citep{terwal}. 
In this latter stream of literature, network analysis tools have been adopted to explore EU-funded collaborative Research \& Development (R\&D) networks, derived from Horizon 2020 and Horizon Europe, key European initiatives supporting research and innovation. 
Data on research topics, objectives, organisations and other details on Horizon projects are available through the open access portal Community Research and Development Information Service (CORDIS) \citep{EuropeanCommissionCordis:Results}.

From the CORDIS dataset, scientific collaboration networks can be constructed, where vertices represent participating organisations, and links correspond to EU-funded research projects in which they are involved. By mapping these connections, researchers can analyse how knowledge is transferred, innovations emerge, and technological capabilities develop across different sectors and countries. Such networking is often referred to as the Triple Helix model \citep{etzkowitz1995triple}.

The topology and the community structure of these networks reflect not just scientific interactions, but also broader EU innovation strategies, policy impacts, and technological diversification efforts. In general such studies can be 
Studies examined network structures in relation to EU policy objectives \citep{Breschi2004UnveilingProgrammes}, dynamics of innovation \citep{Balland2019SmartDiversification, Balland2019Network20032017, Maruccia2020EvidenceHelix}, and the importance of technological diversity \citep{Muscio2022TechnologicalPolicy}. 
\cite{Barber2006NetworkProjects} were among the first to analyse a large European R\&D network, focusing on its topological properties. 
However, a limited number of studies have applied community detection techniques to networks derived from such data, explicitly aiming to capture the tendency of specific organisations or regions to cluster together due to one or more underlying aggregation mechanisms. 
Notable exceptions include \cite{barber2013community}, who analysed the bipartite network of organisations and projects funded by the Fifth European Framework Programme using a variation of the label propagation algorithm and identifying relevant substructures characterised by spatially heterogeneous community groups. 
More recently, \cite{genova2024analytic}, starting from a multipartite network of projects and using a community detection algorithm, detected field-specific collaboration clusters where some European countries play a central role in the project participation. Another study, \citep{KOSZTYAN2024124417} introduces a novel link prediction model for analysing Horizon 2020 collaboration networks, while also identifying communities and assessing their regional distribution. Although not explicitly a community detection method but rather a clustering approach, \cite{cerqueti2024clustering} cluster yearly EU-funded research collaboration networks by combining nodal centrality measures with rank–size analyses, enabling comparisons across years based on the roles of institutions and highlighting both similarities and differences in leading hubs and overall network structure.
The presence of attributes, including scientific disciplines and project topics, guided the identification of distinct thematic sub-networks. \cite{Morea2024MappingAnalysis} and \cite{Morea} analysed specific thematic Horizon networks through centrality measures and community detection.

%On the one hand, the degree distribution of knowledge networks captures the level of hierarchy within networks. It gives a first measure of the ability of networked organisations to coordinate their actions. On the other hand, the degree correlation captures the level of assortativity of networks. It gives a measure of the ability of knowledge to flow between highly and poorly connected organisations.
%We are interested in how the research focus and the connectivity of nodes influence the flow of information and the consolidation of leadership roles in the system. 

All these studies have shown, in the European context, that these networks are crucial for understanding how transnational research collaborations contribute to competitive innovation ecosystems.
The analysis of such data offers comprehensive information on how academic and research institutions, private companies, and public organisations across Europe collaborate to achieve highly innovative outputs.  

The present paper examines EU-funded hydrogen projects, not only for their strategic importance in the energy transition, but because they provide an ideal context for analysing the dynamics through which collaborative innovation networks are formed and structured \citep{Morea}. Hydrogen valleys are integrated regional ecosystems in which research, industry, public authorities, and civil society work together to develop hydrogen-based solutions. These projects are designed to stimulate interaction between heterogeneous actors and promote innovation through a systemic approach. 
Our objective is to analyse the relational structures that develop among organisations involved in Horizon hydrogen projects from 2015 to 2029. Previous analysis on the same dataset demonstrated that the overall knowledge flow is shaped by some pivotal organisations \citep{Morea2024MappingAnalysis}. As such, hydrogen valleys provide an exemplary case for studying leader-influence mechanisms that drive the formation of research communities. 

\subsection{Methodological background}

Community detection in social networks has been extensively studied, with numerous methods developed to address this complex problem \citep{Fortunato2016CommunityGuide, Rosvall2019DifferentDetection,Morea2025SolutionSpace}. 
Most approaches aim to identify groups of nodes that are more densely connected internally than externally; however, they differ substantially in how density is modelled, estimated, or operationalised. 
In addition,  many assume that all nodes in the network hold equal influence, overlooking the presence of leaders or neglecting the availability of attributive information. Recognising the importance of leader influence and homophily mechanisms, few studies have explored their role in defining the community structure. 

Probabilistic models represent communities through a latent partition of the nodes that drives connection patterns. The stochastic block model (SBM) and its extensions \citep{Lee2019} represent the canonical framework, assuming that the probability of an edge between two nodes depends solely on their community memberships. In networks with degree heterogeneity, the degree-corrected SBM \citep[DC-SBM,][]{Karrer2011StochasticNetworks} allows nodes within the same group to have different expected degrees, better reflecting real-world communities where hubs and peripheral nodes coexist.
Hierarchical variants of the SBM explicitly model multi-layered organisation, enabling the detection of hierarchical community structures that arise from nested dense regions surrounded by sparser ones \citep{Peixoto2014HierarchicalModels,Peixoto2019BayesianBlockmodeling,Come2021HierarchicalLikelihood}.
Recent contributions integrate node attributes directly into the probabilistic framework \citep{Stanley2019StochasticAttributes}.

Spectral clustering (SC) and its extensions identify communities by partitioning the network through eigenvectors of the network's Laplacian matrix. Methods such as Covariate-Assisted SC (CASC) \citep{Binkiewicz2017} and Network-Adjusted Covariates (NAC) model \citep{Hu2024Network-adjustedDetection} incorporate node attributes by modifying the similarity matrix or regularising structural information with attribute-based similarities. In particular, NAC aggregates covariate information from neighbours, weighting attributes by connection strength. This adjustment implicitly emphasises high-degree or influential nodes and has been shown to outperform the popular SC and CASC methods.

Density-based approaches interpret communities as dense regions separated by sparser areas, aligning directly with the intuitive definition of community structure. From a statistical perspective, this view is rooted in the modal formulation of clustering, where groups correspond to high-density regions of an underlying distribution, as discussed in \cite{Menardi2015}. Early adaptations such as DENGRAPH \citep{Falkowski2007DENGRAPH:Algorithm} extended DBSCAN \citep{Ester1996} to graph data, identifying communities as locally dense neighbourhoods.

Leader-based methods represent a related but distinct paradigm: they assume that communities form around influential or central nodes, with other nodes gravitating toward these leaders based on structural proximity or attribute similarity. Early algorithms such as LICOD \citep{Yakoubi2014LICOD:Networks}, influence-propagation models \citep{Helal2017LeaderbasedNetworks}, and eigenvector-based leader detection \citep{Ahajjam2018ANetworks} exemplify this idea. 

A key advantage of DeCoDe approaches is that they accommodate heterogeneous cluster shapes and variable node density, thus the presence of multiple dense cores, making them suitable for networks exhibiting such structural patterns. 
An important recent contribution in this direction \citep{Menardi2022Density-basedNetworks} defines leaders as density peaks in the relations space and assigns remaining nodes to their nearest denser neighbour along topological paths. 

Like most leader-based (or leader-first) methods, DeCoDe relies on structural measures to identify influential nodes. However, these measures alone may not fully capture social influence, particularly in networks where node attributes provide additional context. 
To address this, recent leader-based algorithms have incorporated attributive information alongside topology. For example, the aLBCD method \citep{Helal2016AnNetworks} assigns leaders based on attribute similarity when topological data is incomplete, effectively treating attributes as a complement to structural information. Similarly, TALB \citep{Lu2021Leader-BasedNetworks} incorporates attribute information before community detection by modifying the original network structure through a non-fixed weighted topology approach \citep{Chunaev2020CommunitySurvey}.

A key conceptual distinction in the leader identification step concerns the source of node popularity, which can emerge either endogenously--through preferential attachment processes within the network’s structure--or exogenously, from actor attributes such as expertise, complementarity, or availability of resources \citep{Barabsi1999, Bianconi2001, Papadopoulos2012}. Most community detection methods, including DeCoDe, primarily capture endogenous popularity by identifying structurally central nodes that attract many links due to their position. However, in many social and collaborative systems, influence is also shaped by exogenous popularity: nodes become focal points not only because they are well-connected, but because they possess valuable or relevant characteristics that attract others. Accounting for this aspect enables a more realistic representation of community formation mechanisms, where connectivity emerges jointly from structural and attribute-driven affinity.

Probabilistic, spectral, and density-based approaches offer complementary views of community structure: the first rely on generative assumptions, the second on the global geometry of the graph, and the third emphasises local density and heterogeneous node influence. For networks such as Horizon collaborations--characterised by topic-specific subgroups and marked degree variability--the latter perspective is particularly interesting. 

Building on the density-based framework of \cite{Menardi2022Density-basedNetworks}, our approach identifies clusters that emerge around actors located in high-density regions of the attribute space. Under homophily, such representative nodes act as focal points, around which others with comparable attributes tend to cluster. Structural connectivity is preserved by requiring paths in the original network to link nodes, ensuring that communities reflect both attribute similarity and topological integrity.

\section{Methodology}\label{sec:method}

\subsection{Attribute-driven leader identification}\label{subsec:leader_ident}

A central challenge in social network analysis is the identification of influential individuals--often referred to as leaders--who shape the structure and evolution of communities. Traditional approaches to leader detection rely primarily on structural metrics such as degree, betweenness, or eigenvector centrality \citep{Newman2010}. While these measures capture positional importance in the network topology, they overlook a crucial dimension of social interaction: the attributes of individuals. In many real-world networks, homophily plays a fundamental role in shaping community boundaries.

This limitation is particularly evident in large collaborative networks, where structural importance can be misleading. Highly connected nodes often act as bridges across subgroups, causing structural centrality to merge otherwise distinct communities. Conversely, leaders identified through local density typically correspond to small, near-clique subgraphs that may merely reflect project-level co-participation \citep{Menardi2022Density-basedNetworks}.

Here, we propose an attribute-driven perspective on leader identification. Leaders are characterised by high density in the attribute space. This means that they are surrounded by a concentration of nodes sharing similar attribute values, thereby serving as focal points around which communities form.
This attribute-driven leader identification enables the detection of leaders who are not primarily structurally important, but rather attribute-representative. This enhances the interpretability of detected communities by anchoring them around actors whose attributes reflect the dominant features of their groups. 

\begin{figure}[t!] 
    \centering 
    
    % First subfigure
    \begin{subfigure}[t]{0.35\textwidth}
        \centering
        \includegraphics[width=\textwidth]{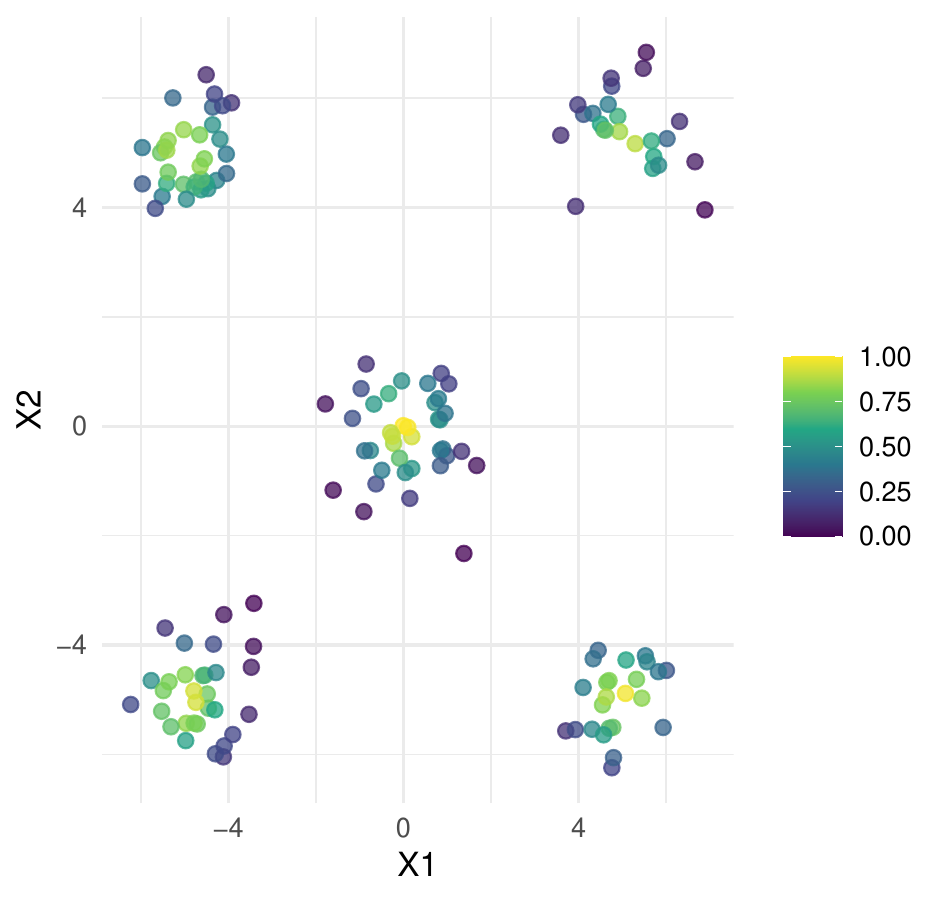}
        \caption{}
        \label{fig:sub1}
    \end{subfigure}
    \hfill
    % Second subfigure
    \begin{subfigure}[t]{0.3\textwidth}
        \centering
        \includegraphics[width=\textwidth]{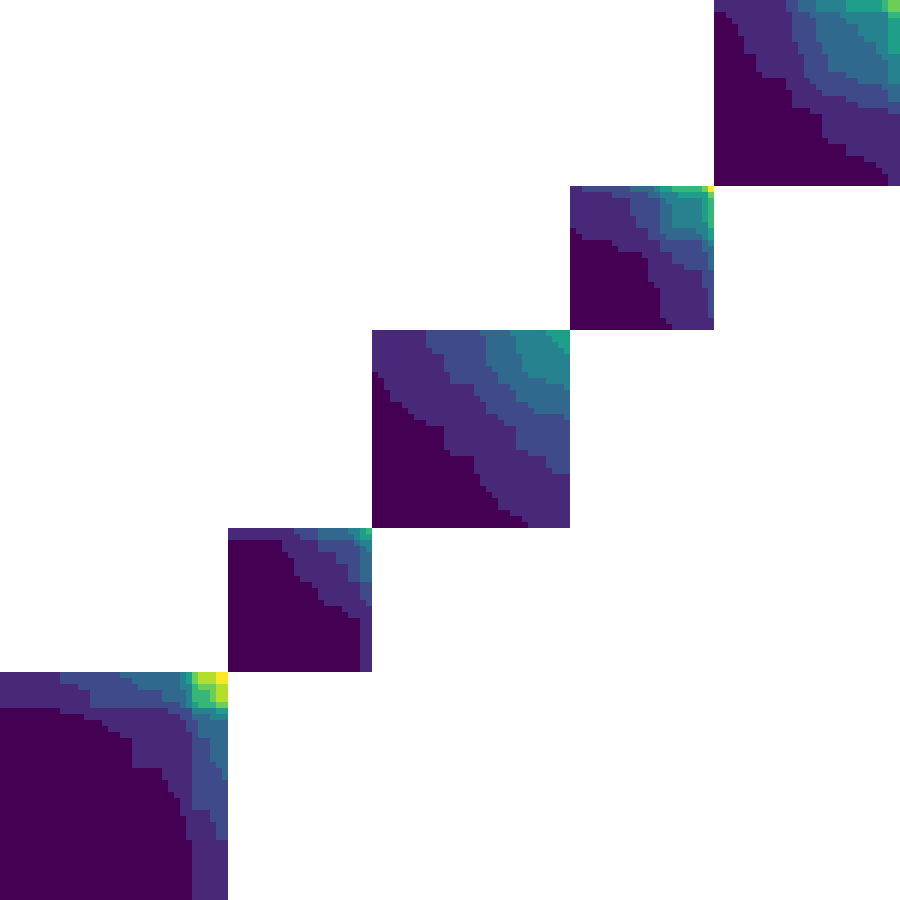}
        \caption{}
        \label{fig:sub2}
    \end{subfigure}
    \hfill
    % Third subfigure
    \begin{subfigure}[t]{0.3\textwidth}
        \centering
        \includegraphics[width=\textwidth]{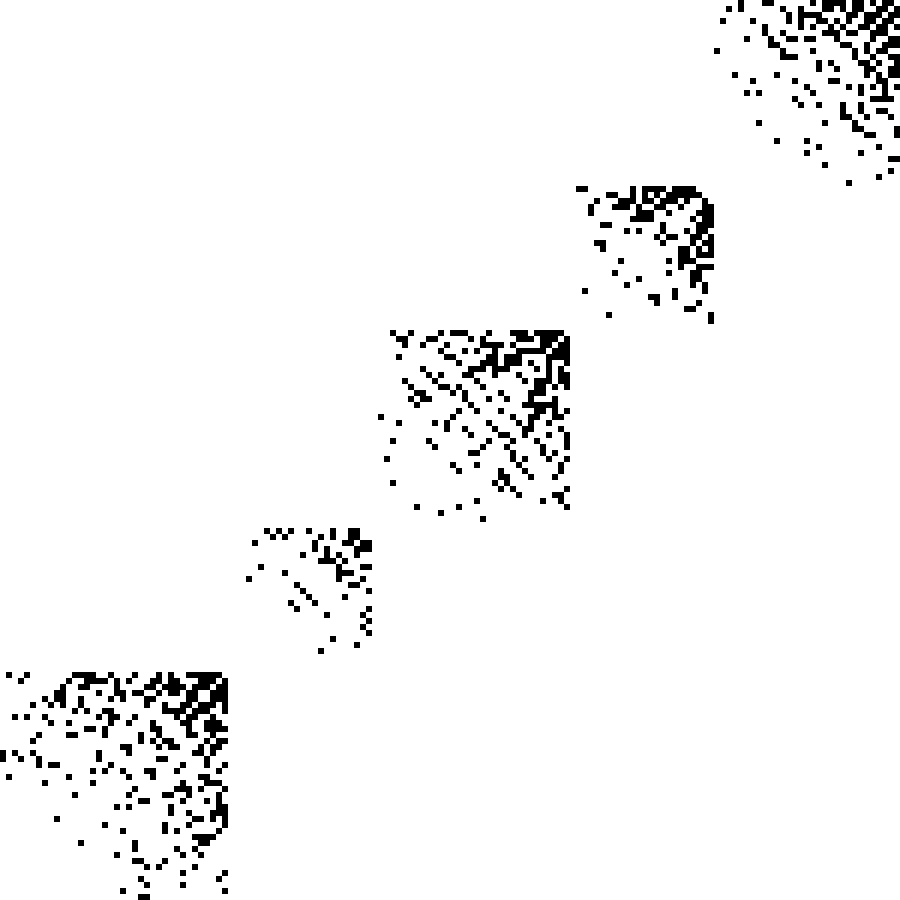}
        \caption{}
        \label{fig:sub3}
    \end{subfigure}

    \caption{(a) Data points generated using a Gaussian model with 5 components, with colour representing density;  (b) Block-diagonal edge probability matrix, with assortative-only blocks representing communities; (c) Block-diagonal adjacency matrix, with assortative-only blocks representing communities.}
    \label{fig:fig1}
\end{figure}

Figure \ref{fig:fig1} illustrates the intuition.
Panel (a) displays five well-separated clusters of data points, with colour representing local attribute density. Within each cloud, density varies, and only a few points (highlighted in yellow) lie at the centre of their group, acting as attribute-representative nodes. These nodes constitute ideal leaders in scenarios where similarity of attributes drives the emergence of communities.
The central panel (b) translates this configuration into a homophily-consistent connectivity pattern, represented in a block-diagonal edge-probability matrix. Nodes that exhibit high attribute density are assigned higher connection probabilities to neighbouring nodes. Within each block nodes are ordered by expected connectivity, with high-density leaders in yellow at the top of each diagonal block. 
Panel (c) shows the corresponding adjacency matrix, generated according to the probabilities in (b). The resulting network exhibits five assortative-only communities, each organised around its attribute-defined leader.

Although simplified for illustrative purposes, this example captures a structure in multiple attribute-coherent and degree-heterogeneous groups of nodes that we expect to observe in collaboration networks such as Horizon. Importantly, even in this idealised assortative setting, community recovery is non-trivial: within each block, most nodes are only sparsely connected, reflecting the heavy-tailed degree distributions typical of real-world data.

To uncover such structure, we extend the DeCoDe framework introduced by \cite{Menardi2022Density-basedNetworks}. Let $G = (V, E, \bm{X})$ denote a node-attributed unweighted network, where 
$V = \{v_1, \dots, v_i, \dots, v_n\}$ is the set of nodes, 
$E = \{ e_{ij} \}$ is the set of (undirected) edges, $i,j = 1, \ldots, n$, $i \neq j$,  with $e_{ij} \in \{0,1\}$ indicating the presence ($1$) or absence ($0$) of an edge between nodes $v_i$ and $v_j$, ; 
$\bm{X} \in \mathbb{R}^{n \times p}$ is the attribute matrix whose $i$-th row $\bm{x}_i$ collects the $p$ attributes associated with node $v_i$.

The DeCoDe approach leverages local density in the network space to identify cluster cores and partition communities. Building on this idea, we integrate attribute information into the density estimation process. Specifically, we estimate the density for node $v_i$ in a continuous attribute space,
$$
\hat\delta(v_i) = \hat f(\bm{x}_i),
$$
where $\hat{f}(\cdot)$ is a suitable probability density function estimator, reflecting the extent to which each node is representative of its neighbours in terms of relevant characteristics. We refer to the resulting method as the {\em Attribute-driven DeCoDe} (AttDeCoDe) approach.

Several alternative methods can be employed to estimate attribute-based densities.
For clarity, we summarise the three approaches adopted in this work:

\begin{enumerate}
    \item \textbf{$k$-Nearest Neighbours (kNN)}
    A node’s density is computed as the inverse of the average distance to its $k$ closest neighbours in the attribute space. This non-parametric estimator measures local attribute concentration without imposing any functional form on the underlying distribution.

    \item \textbf{Gaussian Mixture Models (GMM)}
    A model-based density estimator fitting a mixture of multivariate Gaussian distributions to the attribute space \citep{Bouveyron_Celeux_Murphy_Raftery_2019}. Dense regions correspond to areas of high mixture likelihood. A node’s density is calculated as the weighted sum of component likelihoods, where weights are the estimated mixing proportions.

    \item \textbf{Component-specific multivariate Gaussian densities (GMM-based alternative)}
    Particularly suitable when the attribute space contains well-separated Gaussian-like clusters. 
    This practical alternative first fits a GMM and then assigns each node to the component with the highest posterior probability. The density for the node is evaluated under that component’s multivariate Gaussian distribution, defined by its estimated mean vector and covariance matrix. 
    This procedure yields a collection of component-specific density estimates for each observation, according to its component membership, rather than a single global mixture density estimate.
\end{enumerate}

These approaches offer a flexible way to capture homophily in continuous attributes and can be selected depending on the distribution and dimensionality of the attribute data.

\subsection{Attribute-driven density-based community detection (AttDeCoDe)}

The clustering problem in our context is framed in terms of detecting high-density regions on the network, where nodes are grouped according to a node-wise measure of density \citep{Menardi2022Density-basedNetworks}. Two nodes belong to the 
same cluster if their densities exceed a given threshold and they are connected 
through a path in $G$. This idea extends modal clustering to a relational setting: node-wise densities $\hat\delta(v_1), \ldots, \hat\delta(v_i), \ldots, \hat\delta(v_n)$ are estimated, and communities form by aggregating nodes around modal actors, i.e. local maxima of the density function.

In unweighted networks, for any threshold 
$\lambda$, the upper-level set $V_\lambda = \{ v_i \in V : \hat\delta(v_i) \ge \lambda \}$ 
induces a subgraph $G_\lambda = (V_\lambda, E_\lambda)$, where $E_\lambda = \{ e_{ij} \in E : v_i, v_j \in V_\lambda \}$. 
The connected components of $G_\lambda$ represent density-based groups. As 
$\lambda$ decreases, these components merge, producing a cluster tree that captures the hierarchical organisation of communities. The number of network clusters $K$ is not fixed a priori, but is determined by the structure of this tree: clusters are identified as the connected components of $G_{\lambda}$ at the lowest $\lambda$ levels where they appear as leaves of the cluster tree. Low-density nodes may either remain unassigned or be attached to nearby cluster cores. In addition to the detected number of clusters $\hat K$ and the cluster membership $C(v_i) \in \{1, \ldots, K\}$, the algorithm outputs an indicator $r(v_i)$ specifying whether node $v_i$ acts as a cluster core ($r(v_i)=1$) or as a non-core member ($r(v_i)=0$).

Building on this framework, AttDeCoDe (Algorithm~\ref{alg_attdecode}) generalises DeCoDe to node-attributed networks $G = (V, E, \bm{X})$, where node densities are computed from the attribute matrix $\bm{X}$ rather than from network structural properties. Attribute-based densities capture homophily and similarity in the attribute space, while the network topology still constrains the paths through which clusters form. This ensures that communities are both attribute-coherent and structurally feasible. AttDeCoDe therefore shifts the focus from purely structural density to attribute-informed density, allowing the clustering process to reflect meaningful variation in the attributes that characterise each node. 

\begin{footnotesize}
\begin{algorithm}[b!]
\caption{AttDeCoDe}
\begin{algorithmic}[1]

\State \textbf{Input:} 
\Statex \quad Attributed network $G = (V,E,\bm{X})$ with $n$ nodes
\Statex \quad Attribute-density estimator $\hat{f}(\cdot)$, as obtained from the approaches in Section~\ref{subsec:leader_ident}

\vspace{0.2cm}
\State \textbf{Step 1: Attribute-driven leader identification}
\State Compute $\hat\delta(v_i) = \hat{f}(\bm{x}_i)$ where $\bm{x}_i$ is the attribute vector of node $v_i$

\vspace{0.2cm}
\State \textbf{Step 2: Density-based community detection}
\FOR{$0 < \lambda < \max_i \hat\delta(v_i)$}
    \State Determine the upper level set:
    \STATEx \quad $V_{\lambda} = \{ v_i \in V : \hat\delta(v_i) \geq \lambda \}$
    \State Build the induced subgraph:
    \STATEx \quad $G_{\lambda} = (V_{\lambda}, E_{\lambda})$ where $E_{\lambda} = \{ e_{ij} \in E : v_i, v_j \in V_{\lambda} \}$
    \State Identify the connected components of $G_{\lambda}$
\ENDFOR

\vspace{0.2cm}
\State Build the cluster tree associating each $\lambda$ with the number of connected components of $G_{\lambda}$

\vspace{0.2cm}
\State Identify clusters as the connected components of $G_{\lambda}$ at the lowest $\lambda$ levels (tree leaves)

\vspace{0.2cm}
\State \textbf{Output:}
\Statex \quad Role $r(v_i) \in \{\text{core}, \text{member}\}$
\Statex \quad Cluster membership $C(v_i)  \in \{1, \ldots, K\}$
\Statex \quad Number of detected clusters $\hat K$

\end{algorithmic}
\label{alg_attdecode}
\end{algorithm}
\end{footnotesize}

For completeness, we note that DeCoDe is recovered as a special case of AttDeCoDe. If the attribute-based density in Step~1 of Algorithm~\ref{alg_attdecode} is replaced with a structural density estimator based on $G$--such as degree, 
betweenness centrality, or other topology-derived measures--one obtains the original approach. DeCoDe is therefore a simpler variant that does not account for attribute information and relies solely on topology to define dense regions.

Nonetheless, estimating density from the feature matrix does not preclude 
structural information from influencing the identification of modal actors. 
In the context of networks constructed from bipartite data, such as organisation--project or author--paper networks, event-level attributes are transferred to nodes through projection. The resulting feature matrix encodes relational patterns: nodes participating in many events, or in events characterised by similar attribute profiles, naturally accumulate higher attribute density. 
Consequently, modal actors may emerge from a combination of attribute similarity and structural prominence, even though density is formally computed only from attributes. This mechanism is central to our application, where project-level textual descriptors shape each organisation’s attribute representation and thus influence the density landscape from which modal actors are identified.
More generally, alternative feature-construction strategies—such as node representations that explicitly combine structural prominence with attribute information—could be adopted to encode both dimensions jointly; however, exploring these extensions lies beyond the scope of the present work.

%When weights are present, the same logic applies, but the strength of ties must be explicitly accounted for. In a weighted network, $w_{ij}$ expresses the intensity of the relationship between nodes $i$ and $j$. The clustering process is refined by considering that two nodes are grouped only if they are both dense and strongly connected.

%To operationalise this idea, two alternative aggregation rules can be applied:

%\begin{itemize}
%    \item AND option: two nodes are joined only if each represents the other’s strongest remaining connection. This stricter condition tends to yield a larger number of highly homogeneous clusters.
%    \item OR option: nodes are aggregated when the connection is the strongest for at least one of them, leading to broader, leader-driven clusters where high-density nodes attract their less dense neighbours.
%\end{itemize}

%In both cases, as $\lambda$ decreases, clusters expand by incorporating less dense nodes connected via increasingly weaker ties. 

%From a computational standpoint, the algorithm requires $O((n + m)n)$ operations on a binary network, plus the cost of computing node densities, which depends on the chosen measure. Kernel density estimation (KDE) and k-nearest neighbours (kNN) scale efficiently to thousands of nodes, while Gaussian mixture models (GMM) involve a higher computational cost but offer a probabilistic interpretation of dense regions in the attribute space.

\section{Simulation study}\label{sec:simulation}

\subsection{Simulation design}

To evaluate the empirical performance of the proposed method, we designed a simulation study based on a new generative framework that extends the DC-SBM. While the SBM is widely used to generate networks with community structures, the DC-SBM adds the flexibility to capture degree heterogeneity, thus the presence of leaders and less central nodes gravitating through them. Building on this foundation, our framework incorporates attribute-driven leader influence, allowing us to generate networks in which communities arise from both structural and attribute-based mechanisms.

The generative process proceeds in two stages. In the first stage, node-level attributes are sampled from a GMM using $M = \{5, 10\}$ mixture components.  Different network sizes $n = \{50, 150, 300 \}$ are considered. Community membership is determined according to a distribution, which is either uniform (yielding equal-sized communities) or drawn from a Dirichlet distribution (introducing cluster size imbalance). Each Gaussian component is defined by a fixed mean vector in $\mathbb{R}^2$ and a spherical covariance matrix ensuring well-separated clusters. 
Node densities are defined as their evaluation under the resulting Gaussian mixture density, so that higher density values correspond to nodes lying in more populated regions of the attribute space.

In the second stage, the network topology is generated using a DC-SBM with a community structure that matches the Gaussian data cluster structure, and with $K = M$. Within each community, nodes are connected with probability proportional to the product of densities $\delta_i$. These $\delta$ values are derived from Gaussian densities and transformed to amplify differences between high- and low-degree nodes, thereby reproducing degree heterogeneity observed in real-world data. A detailed description of the construction of the $\delta$ values is provided in Appendix~\ref{sec:appendix:sim_details}.

\begin{figure}[t!]
    \centering
    
    % First subfigure
    \begin{subfigure}[t]{0.3\textwidth}
        \centering
        \includegraphics[width=\textwidth]{img-horizon/ex-adjacency_bw.pdf}
        \caption{}
        \label{fig:sub4}
    \end{subfigure}
    \hfill
    % Second subfigure
    \begin{subfigure}[t]{0.3\textwidth}
        \centering
        \includegraphics[width=\textwidth]{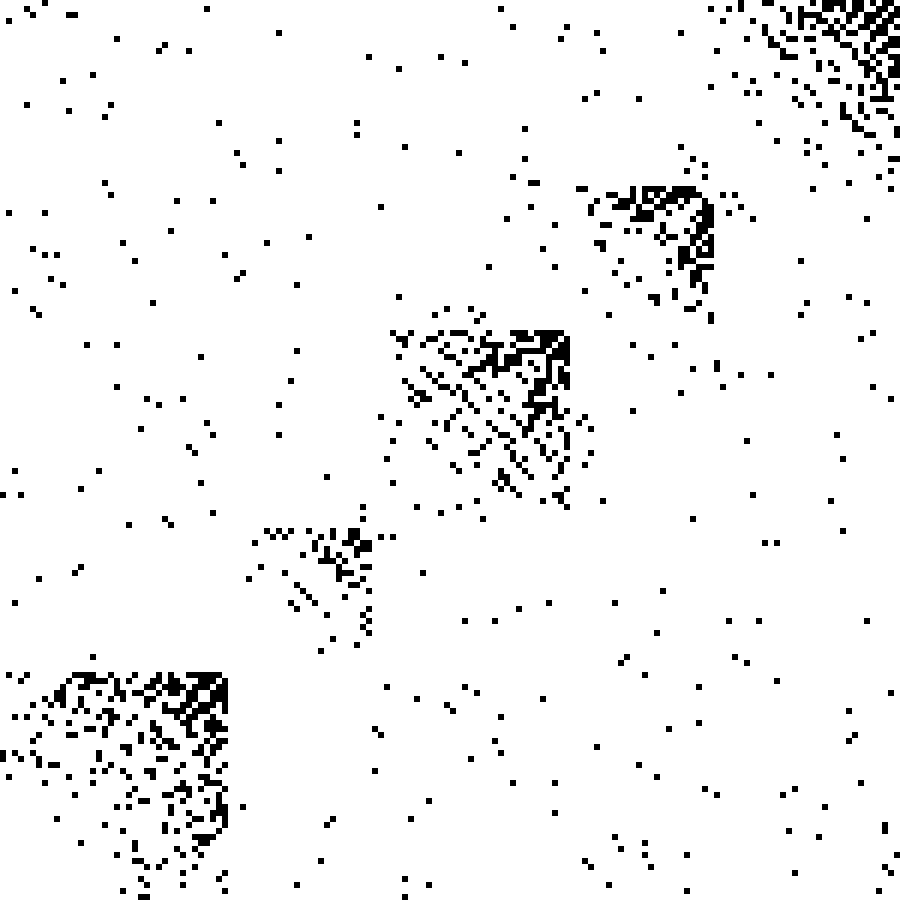}
        \caption{}
        \label{fig:sub5}
    \end{subfigure}
    \hfill
    % Third subfigure
    \begin{subfigure}[t]{0.3\textwidth}
        \centering
        \includegraphics[width=\textwidth]{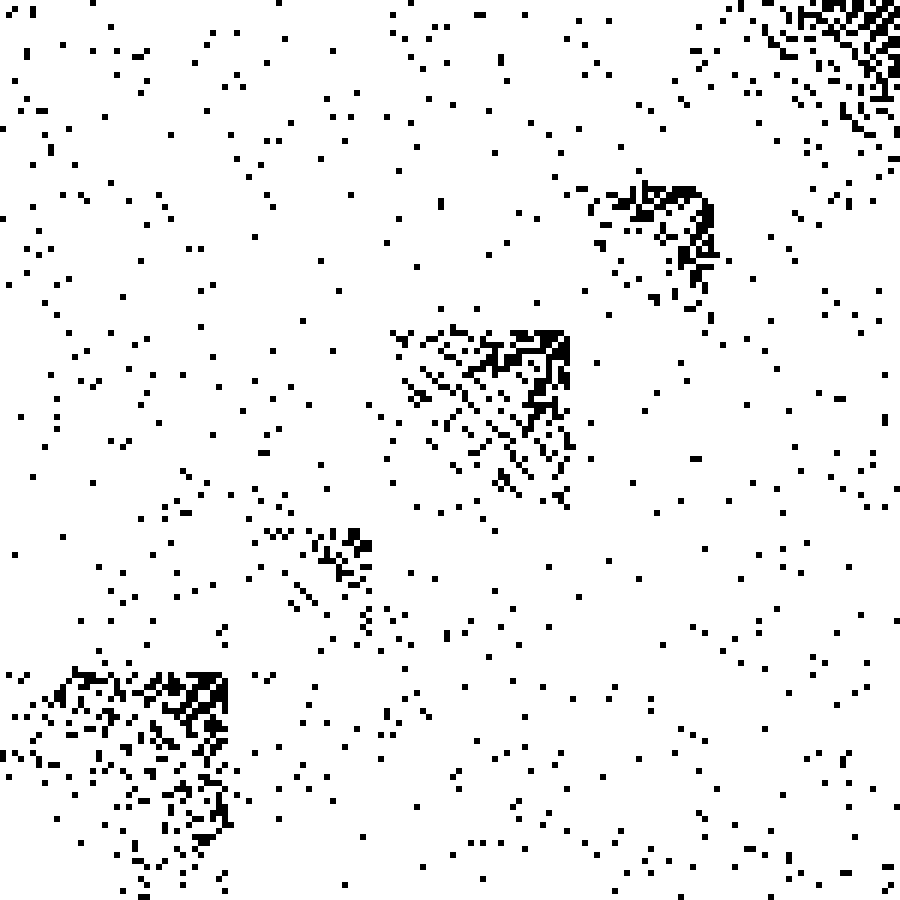}
        \caption{}
        \label{fig:sub6}
    \end{subfigure}

    \caption{Block-diagonal adjacency matrix representing a leader-based network partition in 5 communities; (a) assortative-only, 0\% mixing;  (b)  20\% mixing; (c) 40\% mixing.}
    \label{fig:fig2}
\end{figure}

Cross-community edges are introduced through a controlled mixing procedure, where a proportion of inter-community links are added.  This mechanism is visible in the block-diagonal adjacency matrices displayed in Figure \ref{fig:fig2}: from left to right, the originally ideal assortative structure becomes increasingly “blurred” as dark off-diagonal cells appear, producing a more realistic network topology.

Mixing is governed by a global parameter, $\mu$, which regulates the extent to which nodes connect outside their own community. Increasing $\mu$ leads to a higher proportion of inter-community links and therefore to stronger overlapping between clusters. Namely, $\mu$ represents the ratio between a node’s external links $d_i^{ext}$ and its total degree $d_i$. Thus, when mixing is introduced, a fraction of $\mu$ links connects nodes in different communities \citep{Lancichinetti2008BenchmarkAlgorithms}.
 
To better reflect within-community preferential attachment, we allow mixing rates to vary across nodes. Specifically, we assign each node a weight inversely proportional to node degree:
$$
w_i = \frac{\overline{d}_{C(v_i)}}{d_i},
$$
where $\overline{d}_{C(v_i)}$ is the mean degree in the community of node $i$. These weights are normalised to $[0,1]$ and rescaled so that the expected mean mixing equals the target mixing value $\mu$:
$$
\mu_i = \min\!\left(\tilde{w}_i \cdot \frac{\mu}{\tfrac{1}{n}\sum_j \tilde{w}_j}, \, 1 \right),
$$
with truncation at 1 to avoid invalid probabilities. The number of inter-community connections of each node is then given by $d_i^{ext} = \mu_i \cdot d_i$.

This design yields networks with realistic structural properties: communities exhibit heterogeneous density distributions driven by node attributes, while the level of inter-community mixing is regulated by $\mu = \{0, .2, .4\}$. 
By combining attribute-driven leader attraction with structural connectivity, our simulation framework provides a principled basis for assessing whether the proposed algorithm can recover the true partitioning of the network.

The proposed model is compared against several baseline methods.
Specifically, we consider original DeCoDe, SBM, SC, CASC, and NAC. These approaches are commonly used in the literature and serve as representative competitors for both network-based and attribute-assisted clustering. The binary SBM is implemented via the \texttt{blockmodels} package \citep{blockmodels}, while SC, CASC, and NAC are available in the \texttt{NAC} R package \citep{NAC}.
For the SBM, the number of communities is selected by maximising the Integrated Classification Likelihood (ICL) \citep{Come2015ModelLikelihood}. We then use this ICL-optimal number of components for SC, CASC, and NAC. This choice allows us to anchor the comparison to a statistically principled criterion, rather than tuning $K$ separately as required by SC methods.

The DeCoDe model provides a direct baseline for assessing the impact of our proposed extension. In our comparison, we evaluate AttDeCoDe under two main density specifications:
(i) using the true density function employed to generate the synthetic networks, and
(ii) using empirically estimated densities derived from the observed node attributes. For the empirical specification, we consider three alternative estimators. The first, referred to as the component-wise Gaussian density, computes separate densities for each Gaussian component identified by the fitted mixture model. 
The second estimator is the GMM density, obtained directly from the mixture model implemented via the \texttt{mclust} package \citep{mclust}. 
Finally, we include a nonparametric baseline based on kNN density estimation with $k = 5$, implemented using the \texttt{FNN} package \citep{FNN}.

Clustering performance is assessed by measuring the agreement between the estimated and the true partitions, expressed through the Normalised Mutual Information (NMI) \citep{Danon2005ComparingIdentification}, which ranges from 0 (no agreement) to 1 (perfect agreement). We also compute the Adjusted Rand Index (ARI) \citep{hubert1985comparing}; however, simulation study results based on ARI are not reported here, as they lead to conclusions fully consistent with those obtained using NMI and do not provide additional insights in this setting.

\subsection{Simulation results}

Among the empirical density estimators considered, we adopt the component-wise Gaussian density as the primary specification for our simulation study. This choice is motivated by the design of our generative model, where node attributes are drawn from distinct and well-separated Gaussian components. The GMM density tends to smooth across component boundaries, potentially obscuring inter-cluster separation, while the kNN estimator, though flexible, is less stable in low-dimensional settings with limited sample size.
A detailed comparison of the performance of all three estimation approaches with the true density is provided in the Appendix (Figures \ref{appx:fig1}).

 Figure~\ref{fig:fig3} shows the results of a study that investigates how network size and number of communities influence the ability of our AttDeCoDe approach to detect community structures in networks with equally sized communities. Results for networks with non-uniform community sizes are reported in the Appendix (Figure~\ref{appx:fig2}). Results suggest that our method performs robustly across a range of realistic network configurations.
Because the DeCoDe approach is density-based, its performance is affected by the extent to which leaders form connections with high-density nodes in other communities. To evaluate this behaviour, we evaluate the performance of the detection methods applied to networks with varying proportions of inter-community links (mixing level $\mu$).

Across scenarios for smaller networks ($n = 50$), both versions of the proposed method--AttDeCoDe using the true density ($\delta$) and AttDeCoDe using the empirically estimated density ($\hat\delta$)--consistently outperform the baseline methods in terms of NMI. The improvement over the original DeCoDe is most pronounced when mixing is introduced, where attribute information substantially enhances community separability. As the mixing increases, all methods show a decline in performance due to reduced structural distinctiveness among communities, but AttDeCoDe maintains the highest stability across the full range of mixing.

Increasing network size improves performance for baseline algorithms, indicating better recovery of both structural and attribute-based patterns in larger samples. In contrast, AttDeCode performance remains stable across network sizes. The effect of the number of components in the attribute space is also visible: for $M = 10$, the nodes are grouped in smaller communities, and the network is more fragmented, especially for smaller networks, making community detection more challenging. In this setting, the gain from explicitly modelling density heterogeneity (as in AttDeCoDe) remains evident, whereas simpler structural approaches exhibit a decline in performance.

Overall, these results confirm that AttDeCoDe effectively integrates attribute-driven density information, yielding robust community recovery even under moderate inter-community mixing. 

\begin{figure}[t!]
    \centering
        \centering
        \includegraphics[width=\textwidth]{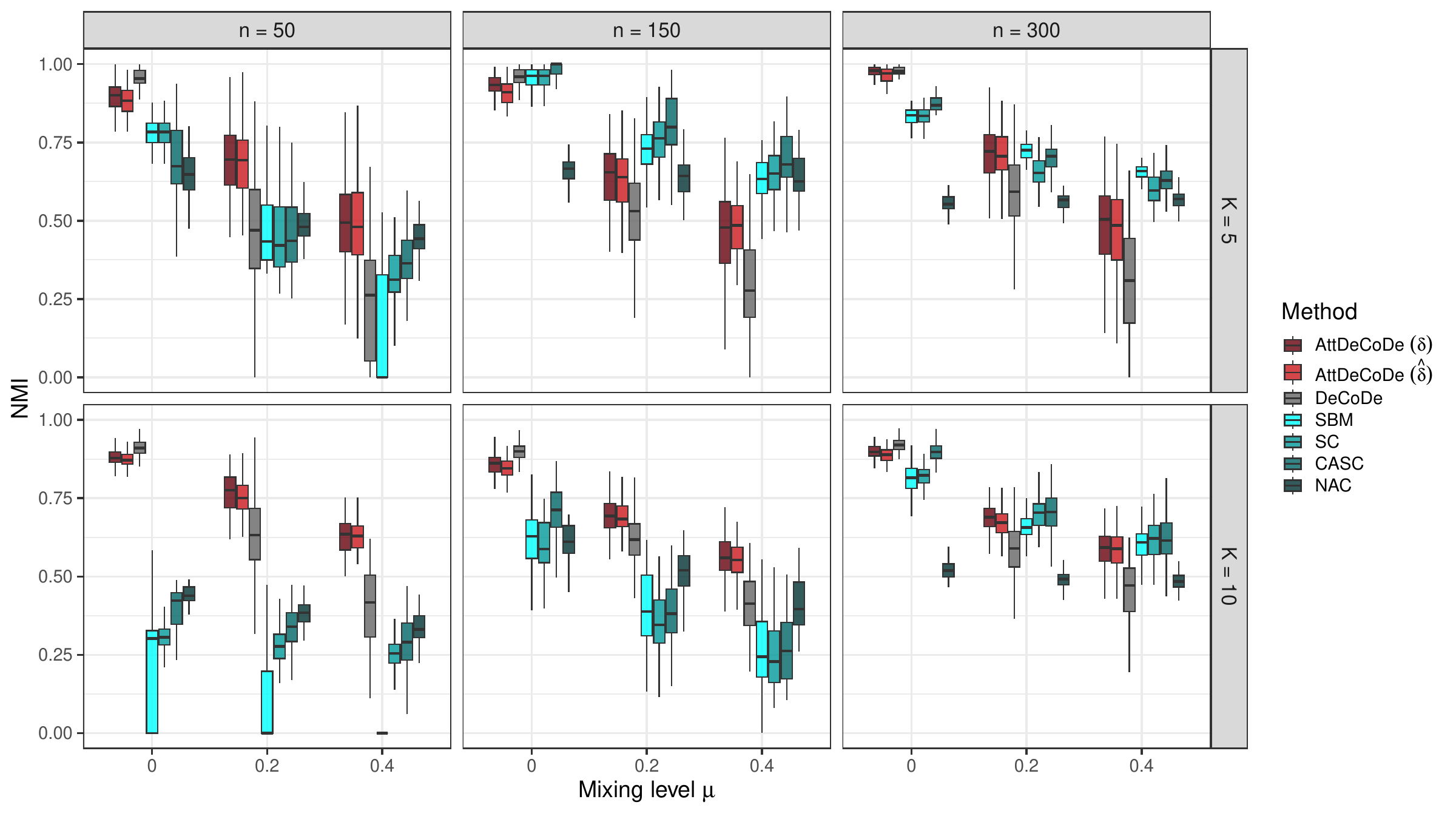}
        \caption{Normalised Mutual Information (NMI) distribution across different network sizes and numbers of communities ($K$) for \textbf{uniform} community sizes. Results are reported for all competing methods, including AttDeCoDe, DeCoDe, SBM, Spectral Clustering (SC), Covariate-Assisted SC (CASC), and SC on Network-Adjusted Covariates (NAC).}

    \label{fig:fig3}
\end{figure}

\subsection{Benchmark network results}

\begin{figure}[t!]
    \centering
        \centering
        \includegraphics[width=\textwidth]{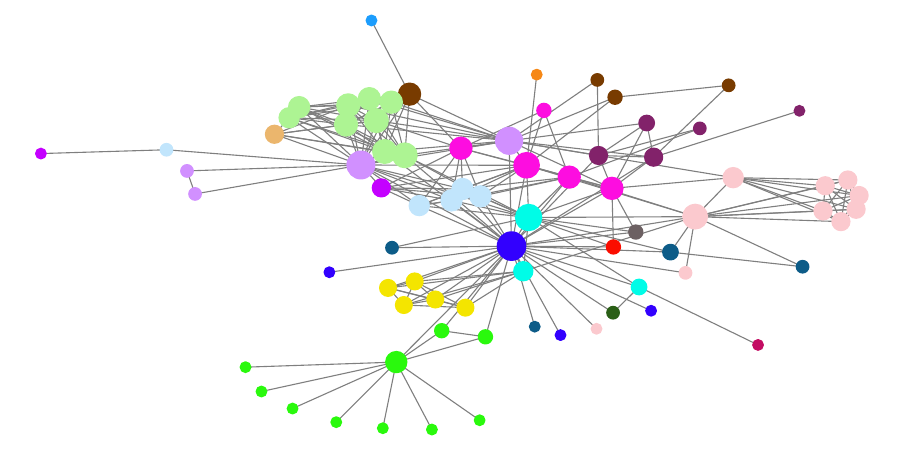}
        \caption{}

    \caption{Les Misérables network partition in 20 communities with point size representing node degree and node colour indicating community membership.}
    \label{fig:fig4}
\end{figure}

The evaluation of the detection performance involves examining how our method performs on a popular dataset, where community membership and attributes of nodes are available. 
The popular Les Misérables character network (Figure \ref{fig:fig4}) describes the interactions between 77 characters in Victor Hugo's novel Les Misérables \citep{KnuthTheComputing}. The edges represent the co-appearance of characters in one or more chapters of the novel. The ground truth membership corresponds to the early appearance of each character in the book, resulting in $K = 20$ small communities.
We used textual descriptions of characters as node attributes, which are accessible from the dataset provided in \citep{Kalugin2015LesData}. We transformed the character descriptions into a Term Frequency-Inverse Document Frequency (TF-IDF) matrix. This numerical representation allows us to incorporate the textual information on the characters into our analysis, enabling a richer understanding of the network.

\begin{figure}[t!]
    \centering
    
    % Second subfigure
    \begin{subfigure}[t]{0.48\textwidth}
        \centering
        \includegraphics[width=\textwidth]{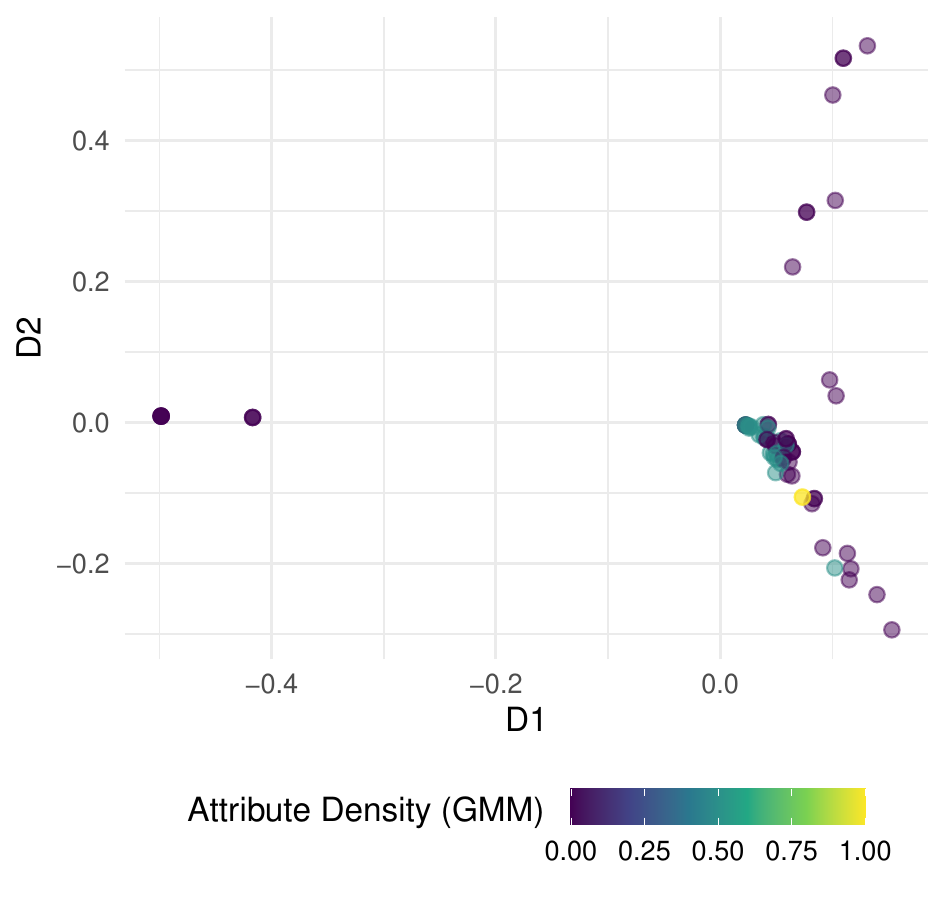}
        \caption{}
        \label{fig:sub7}
    \end{subfigure}
    \hfill
    % Third subfigure
    \begin{subfigure}[t]{0.48\textwidth}
        \centering
        \includegraphics[width=\textwidth]{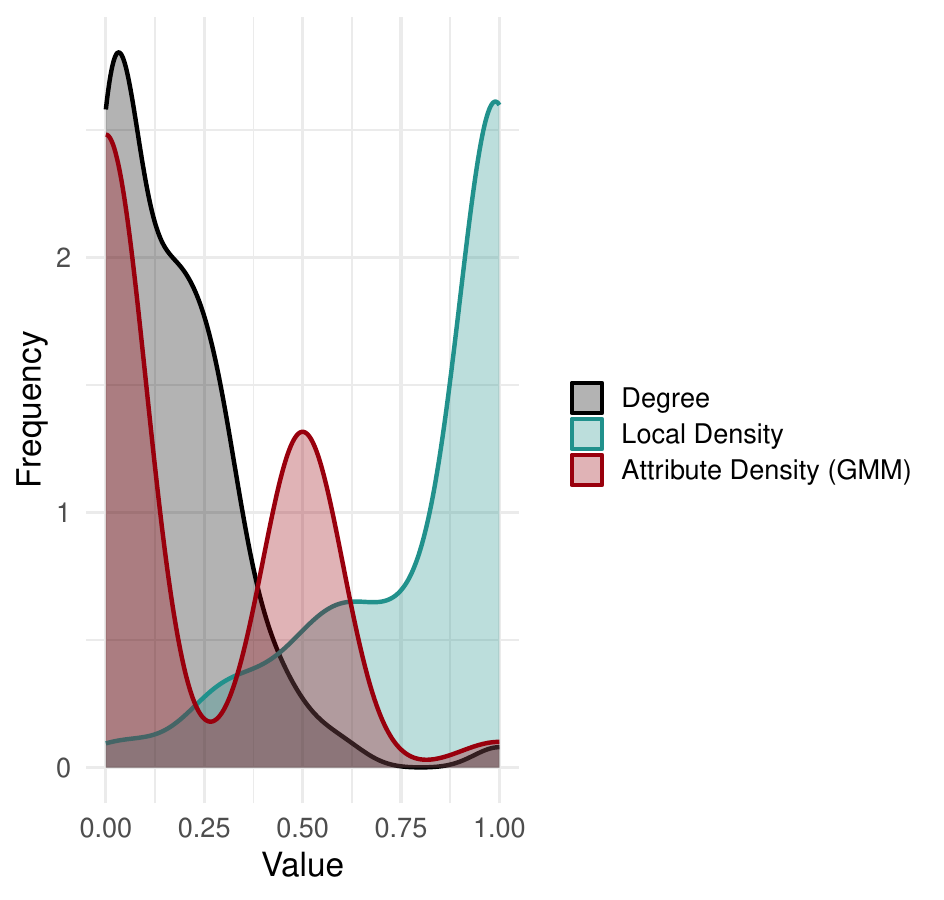}
        \caption{}
        \label{fig:sub8}
    \end{subfigure}

    \caption{Density distribution of nodes in the Les Misérables network (a) 2D MDS projection of node attributes with lighter colours indicating higher GMM density; (b) normalised distributions of GMM density, node degree and local density.}
    \label{fig:fig5}
\end{figure}

For estimating the attribute-based density, we fit a GMM with a large number of components, namely $M = n/2$.
This approach is supported by theoretical findings that a GMM can approximate any smooth density function with arbitrary accuracy, given a sufficient number of components \citep{DeepLearning, Nguyen01012020}. Additionally, this choice is preferable for this dataset, as the small number of data points would lead to poor estimates using methods like kNN.
To visualise the attributive information, we applied multidimensional scaling (MDS) to the cosine distance matrix computed from the TF-IDF document-term matrix. In the resulting 2D embedding (Figure \ref{fig:sub7}), nodes with higher GMM density values are represented with lighter colours (i.e., yellow and green) and tend to appear in the denser regions of the plot. The distribution of the GMM density values differs from that of the node degree or the local density used in the original Decode paper; all quantities are normalised to ensure comparability. As the degree, the GMM density distribution exhibits a peak at low values, but it also shows a secondary local peak around 0.5 (Figure \ref{fig:sub8}).

Table \ref{tab:tab1} presents clustering performance across different methods. 
Among the DeCoDe variants, using node degree alone performs poorly, failing to recover meaningful community structure (NMI $ = 0$, ARI $=0$, $\hat{K} = 1$). Incorporating local density markedly improves performance (NMI $= 0.76$, ARI $= 0.40$, $\hat{K} = 35$), reflecting the benefit of considering neighbourhood density for identifying communities. The proposed method, AttDeCoDe, recovers 22 clusters, which is close to the true number, indicating a smaller degree of over-segmentation than the local density variant. AttDeCoDe achieves the highest agreement with the ground truth among all tested methods, with NMI $= 0.78$ and ARI $= 0.50$, outperforming other standard methods such as SBM, SC, CASC and NAC. This suggests that incorporating attribute-driven density information into the community detection process improves the recovery of the underlying community structure.

\begin{table}[b!]
\centering
\begin{tabular}{lrrr}
  \hline
  Method & NMI & ARI & $\hat{K}$ \\ 
  \hline
 DeCoDe (Degree) & 0.00 & 0.00 &   1 \\ 
   DeCoDe (Local Density) & 0.76 & 0.40 &  35 \\ 
   AttDeCoDe & \textbf{0.78} & 0.50 &  \textbf{22} \\ 
   SBM & 0.20 & 0.54 &   6 \\ 
   SC & 0.32 & 0.63 &   6 \\ 
   CASC & 0.02 & 0.30 &   6 \\ 
   NAC & 0.24 & 0.54 &   6 \\ 
   \hline
\end{tabular}
\caption{Agreement in terms of Normalised Mutual Information (NMI) and Adjusted Rand Index (ARI) between estimated and early appearance partition of Les Miserables network, along with estimated number of clusters ($\hat{K}$), across community detection methods.}
\label{tab:tab1}
\end{table}

\clearpage

\section{Application to the Hydrogen Horizon network}\label{sec:application}

The Hydrogen Horizon network describes the interactions between 859 organisations collaborating on 181 EU-funded projects, yielding a sparse one-mode projection with a density of 0.02\%. For this study, we focus on the largest connected component, which contains 825 organisations. The 7,807 edges in the organisation one-mode network represent relationships derived from the shared connections in the project-organisation bipartite network.
An edge between two organisations is formed if both are involved in at least one common project; the edges can carry weights that reflect the number of shared projects between two organisations. In this setting, edge weights are highly heterogeneous but also strongly driven by project size and consortia composition rules, which can inflate ties of large, central organisations that participate in many consortia. As a result, high weights often reflect prolific participation or administrative roles rather than meaningful leadership or homophilous attraction. Using such weights directly in a density-based clustering framework may bias the identification of clusters by obscuring attribute-driven cohesion. For this reason, in this work we do not interpret edge weights as indicators of stronger connection paths in the clustering process.

Available information includes additional organisation-level and project-level attributes, i.e., geographical location, activity type, and thematic focus. This allows us to investigate whether geographical or cultural proximity, shared research interests, and institutional roles (research institutes, universities, industry, etc.) shape collaboration patterns, leading to the formation of subgroups by characteristics. At the same time, it is worth mentioning that Horizon programme rules encourage diversity within consortia: funded collaborations are expected to include partners from different countries and to promote interaction across activity types, particularly between academia and industry. 

Table~\ref{tab:tab2} reports the distribution of organisations by country and activity type. From the country perspective, the largest shares come from France (13.2\%), Italy (12.1\%), and Germany (10.9\%), followed by Spain and the Netherlands. These countries represent major hubs within the European research area and coordinate a substantial proportion of Horizon-funded projects. Northern and Eastern European countries appear with smaller proportions, while the category RoE (Rest of Europe) aggregates several countries with individually small contributions. Overall, the distribution reflects the concentration of research capacity and Horizon participation in Western and Southern Europe.

With respect to activity type, private for-profit organisations (PRC) dominate the network (62.4\%). Higher education institutions (HES) and research organisations (REC) each account for approximately 13\% of participants, forming the academic and scientific core of the network, while public organisations (PUB) and other entities (OTH) contribute less. This composition underscores the heterogeneous, multi-actor nature of Horizon consortia, which integrate universities, industry, and research centres in mission-oriented collaborations. 

\begin{table}[bt]
\centering
\caption{Distribution of organisations involved in Hydrogen Horizon projects by country, activity type, and dominant research topics.}
% latex table generated in R 4.5.0 by xtable 1.8-4 package
% Tue Nov 11 14:10:48 2025
\begin{tabular}{llrrr}
  \toprule
   & Level &   & N & \% \\ 
    \cmidrule{2-2} \cmidrule{4-5} 
 Country &  FR &  & 109 & 13.2 \\
& IT &  & 100 & 12.1 \\
& DE &  &  90 & 10.9 \\
& ES &  &  71 &  8.6 \\
& NL &  &  69 &  8.4 \\
& NO &  &  41 &  5.0 \\
& BE &  &  37 &  4.5 \\
& EL &  &  37 &  4.5 \\
& UK &  &  34 &  4.1 \\
& DK &  &  28 &  3.4 \\
& FI &  &  23 &  2.8 \\
& SE &  &  17 &  2.1 \\
& AT &  &  15 &  1.8 \\
& RoE & & $< 15$ & 18.7 \\
     \cmidrule{2-2} \cmidrule{4-5} 
Activity type & PRC & & 515 & 62.4 \\
& HES &  & 113 & 13.7 \\
& REC &  & 105 & 12.7 \\
& OTH &  &  61 &  7.4 \\
& PUB &  &  28 &  3.4 \\
& Missing & &  3 & 0.4 \\
     \cmidrule{2-2} \cmidrule{4-5} 
Top 10 keywords & fuel cells & & 316 & 38.3 \\
& electrolysis & & 225 & 27.3 \\
& natural gas & & 111 & 13.5 \\
& ecosystem & & 110 & 13.3 \\
& wind power & & 101 & 12.2 \\
& catalysis & & 88 & 10.7 \\
& sustainable economy & & 87 & 10.5 \\
& business models & & 86 & 10.4 \\
& energy conversion & & 67 & 8.1 \\
& coating and films & & 58 & 7 \\
\hline
\end{tabular}
\label{tab:tab2}
\end{table}

To better understand how research interests shape collaborations, we complement organisational characteristics with information on the scientific content of the projects. Each funded project is described by a list of keywords encoded using the European Science Vocabulary \citep[EuroSciVoc,][]{Euroscivoc}. We summarise keyword frequencies both at the project level (how often a keyword appears across projects) and at the organisation level (how many organisations are associated with each keyword through their project participation), thus distinguishing research topic prevalence from organisational thematic reach.

At the project level, the most frequent keywords reflect the thematic focus of the EU-funded hydrogen research: \textit{fuel cells} and \textit{electrolysis} (54 projects each), \textit{catalysis} (20 projects), and \textit{natural gas} (17 projects). At the organisation level, keyword frequencies reflect thematic reach. \textit{Fuel cells} and \textit{electrolysis} dominate, involving 38.3\% and 27.3\% of organisations respectively, confirming their central role in the network. \textit{Natural gas}, \textit{ecosystem}, \textit{wind energy} (around 12-14\%) indicate widespread engagement in adjacent energy sectors. The presence of \textit{catalysis}, \textit{sustainable economy}, and \textit{business models} (around 10\%) highlights the spread of socio-economic themes across many organisations, while more specialised areas are limited to smaller subgroups of experts. 

Fifty projects contain one or more keywords that do not appear elsewhere, and only four projects have more than half of their keywords classified as unique. Examples include \textit{simulation software}, \textit{mathematical models}, and \textit{thermodynamics} in project 101053133; \textit{metallurgy} and \textit{thermodynamic engineering} in 101091456; \textit{tissue engineering}, \textit{stem cells}, and \textit{piezoelectrics} in 641640; and \textit{crystallography}, \textit{bioinorganic chemistry}, \textit{quantum computers}, and \textit{enzymes} in 745702. These cases highlight the presence of highly specialised research niches embedded within an otherwise thematically coherent research.

The initial observations regarding edge weights are consistent with the concentration of both attributes and connectivity in the network, which displays a markedly heterogeneous degree distribution characterised by a small number of highly connected hubs and a large majority of peripheral organisations (Figure~\ref{fig:sub10}). These hubs are disproportionately composed of private for-profit companies (PRCs), organisations based in Finland, and actors whose dominant thematic focus is \textit{natural gas}, indicating that structural prominence may be closely associated with institutional roles, national participation patterns, and research domains characterised by lower entry barriers and widespread accessibility within the application context.

Collaborative ties in research networks appear to be shaped by the interplay of thematic priorities and leading organisations. 
Although this mechanism is not directly observable from the network structure alone, it seems to appear as a key aspect of bipartite event-actor networks, where the characteristics of the events that generate ties play a key role in identifying potential leaders.
 
 In our application, project topics naturally encode node attributes: an organisation’s thematic profile is derived from the collection of textual descriptors associated with the projects in which it participates. This representation facilitates the identification of similarity-driven attachment through homophily in the attribute space, consistent with popularity–similarity models of network growth \citep{Papadopoulos2012}. At the same time, the same project-level information also captures structural popularity: organisations involved in many projects accumulate broader and more heterogeneous keyword profiles, making them more likely to occupy central regions of the attribute space and to be representative of a wide range of other actors. In this bipartite setting, similarity and popularity are therefore not independent dimensions but are jointly shaped by the underlying structure.

\subsection{AttDeCoDe results}

Topic-specific clouds tend to form around organisations that dominate particular research areas, whereas actors engaged in more interdisciplinary, emerging, or weakly connected topics may occupy a more dispersed position within the network. Our approach estimates attribute-based node density to identify thematic focal points and then applies DeCoDe to delineate the subgraphs that form around them. This perspective is consistent with the logic of Horizon collaborations: organisations engaged in similar research areas tend to cluster around shared priorities, while differences in expertise, mission, or sector generate boundaries that limit the formation of ties.

To identify clusters driven by homophilous influence, we construct node-level attributes by assigning each organisation the complete set of keywords attached to the hydrogen projects in which it participates. This yields a standardised thematic profile for each organisation, allowing us to assess whether proximity in scientific domains is reflected in the observed clustering patterns. By grounding community detection in project-level attributes, our framework identifies leaders as organisations that are both thematically central--exhibiting high attribute density--and structurally well positioned to attract collaborations across multiple projects, thereby accumulating a broad and diverse keyword profile.

We represent each organisation through a dense semantic embedding of its associated keywords and estimate local concentration in the resulting numerical attribute space.
Specifically, we generate sentence embeddings from the concatenated keyword lists using the all-MiniLM-L6-v2 SentenceTransformer model \citep{HuggingFace2024All-MiniLM-L6-v2}, which yields 384-dimensional vectors in a dense semantic space. This embedding approach allows us to capture higher-order semantic relationships among keywords--beyond simple term co-occurrence, and to represent organisations in a continuous attribute space suitable for density estimation.

\begin{figure}[ht!]
    \centering
    
    % Second subfigure
    \begin{subfigure}[t]{0.48\textwidth}
        \centering
        \includegraphics[width=\textwidth]{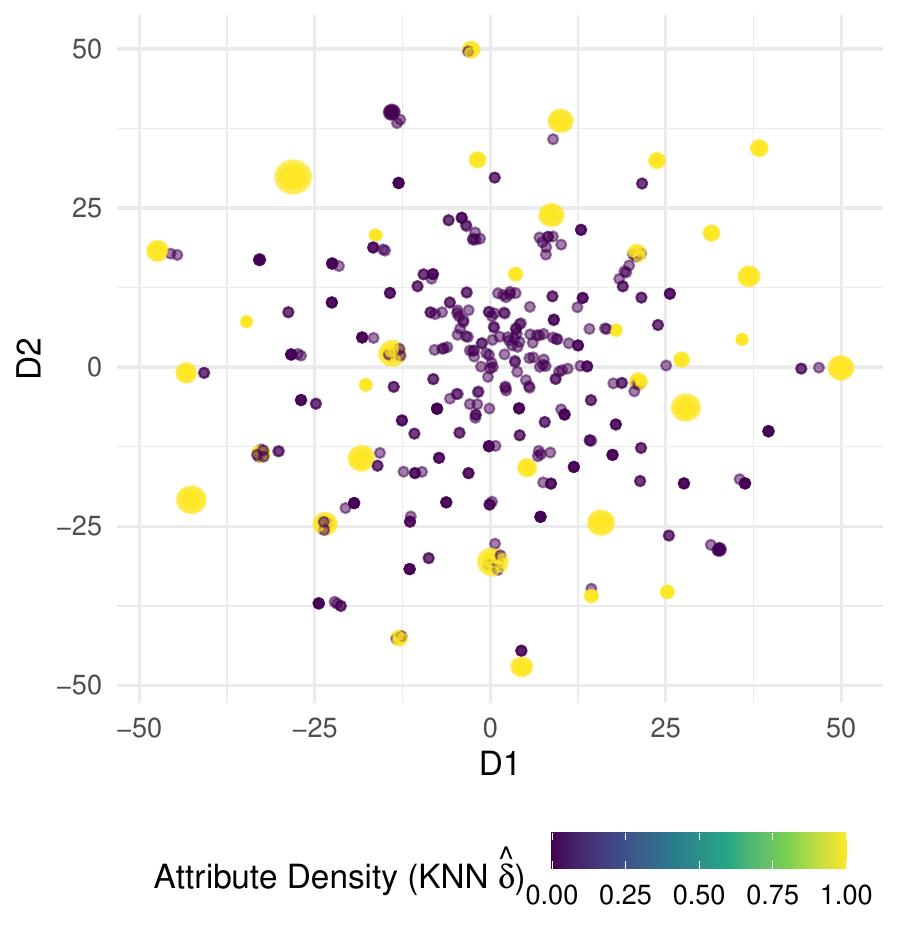}
        \caption{}
        \label{fig:sub9}
    \end{subfigure}
    \hfill
    % Third subfigure
    \begin{subfigure}[t]{0.48\textwidth}
        \centering
        \includegraphics[width=\textwidth]{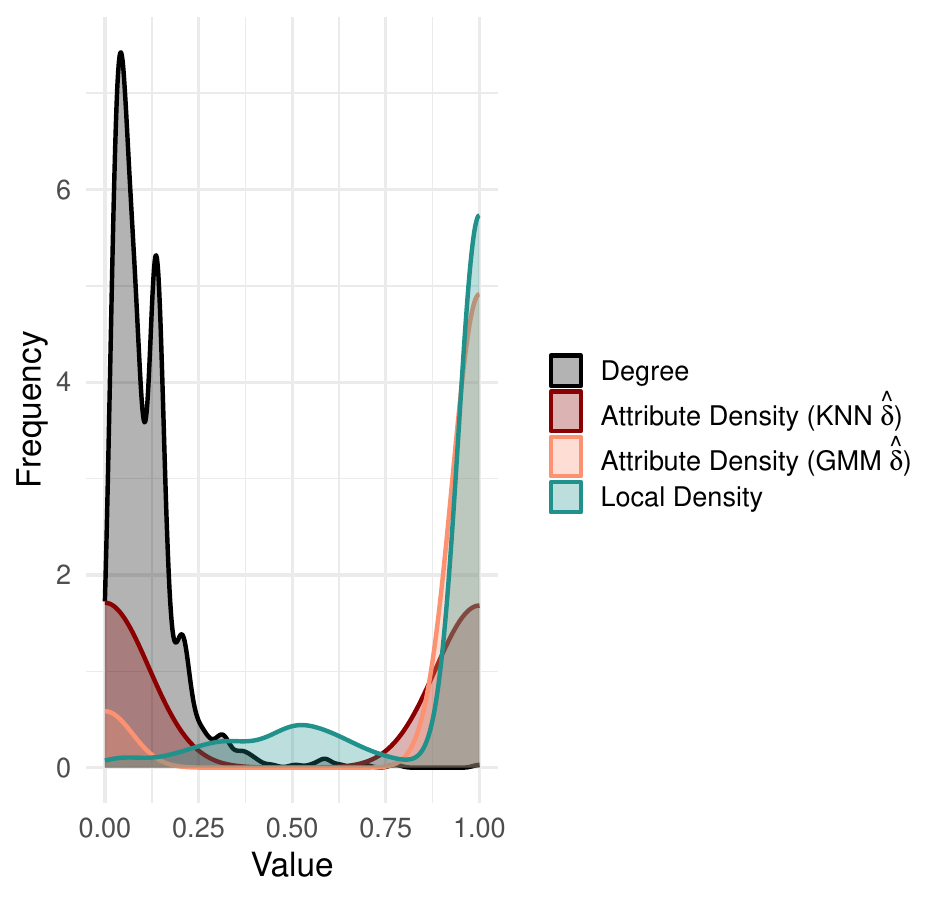}
        \caption{}
        \label{fig:sub10}
    \end{subfigure}

    \caption{Distribution of node density in the Hydrogen Horizon network (a) 2D TSNE projection of node attributes with lighter colours indicating higher kNN density; (b) normalised distributions of kNN density, node degree, and local density.}
    \label{fig:fig7}
\end{figure}

To visualise this high-dimensional attribute data, we apply t-SNE to the embeddings. t-SNE \citep{vanderMaaten2008} is widely used for dimensionality reduction of embedding spaces produced by language models, as it preserves local neighbourhood structure and facilitates the visualisation of underlying clustering patterns.
In the resulting 2D embedding space (Figure \ref{fig:sub9}), attribute-defined dense regions appear as yellow points close in the space. 

For estimating the attribute-based density, we fit a GMM with a relatively large number of components ($M$). Given the high number of points ($n$), using $n/2$ components--as in the benchmark network example--is computationally infeasible, as the Expectation–Maximization algorithm in {\tt mclust} fails to converge due to overparameterization. Instead, we constrained $M$ to be greater than 100 and allowed the optimal number of components to be automatically selected by the Bayesian Information Criterion (BIC).

We also considered a non-parametric kNN density estimator with $k = 5$. This relatively small value of $k$ allows for capturing local variations in attribute concentration while maintaining robustness against noise.

Both GMM and kNN attribute densities capture similar general patterns, being able to identify these visually dense clouds. However, their shapes differ (Figure \ref{fig:sub10}). The GMM estimator assigns the maximum density to approximately 89\% of the nodes, collapsing most of the attribute space into a single high-density mode and failing to distinguish a substantial proportion of low-density points. By contrast, the kNN density assigns high density to around half of the nodes and spreads the remaining ones across lower values, thereby successfully identifying both the localised dense regions visible in the 2D space, as displayed in Figure \ref{fig:sub9}. 

This partition between low- and high-density nodes is not easily captured by structural centrality measures such as degree or local density, which show unimodal distributions--degree concentrated at low values and local density concentrated at high values--providing no clear way to identify meaningful high-density regions. These divergent behaviours also complicate the choice of a centrality measure in the original DeCoDe framework. 
At the same time, the heterogeneous degree distribution of the network, characterised by hubs that span multiple communities, reinforces the need for a density-based approach.
Applied to the Horizon data, DeCoDe using degree produces a trivial solution with a single cluster, offering no analytical insight, whereas DeCoDe using local density heavily over-segments the space, partitioning the 825 organisations into 115 clusters with an average size of only seven nodes.

Instead of relying on structural prominence alone, modelling density in the attribute space provides an appropriate means of uncovering community structure driven by thematic proximity. This is particularly relevant in the Horizon network, where the attractiveness of cluster representatives--and the formation of cohesive groups around them--is plausibly driven by similar research interests rather than by purely structural prominence.

\begin{figure}[t!]
    \centering
    
    % First subfigure
    \begin{subfigure}[t]{\textwidth}
        \centering
        \includegraphics[width=\textwidth]{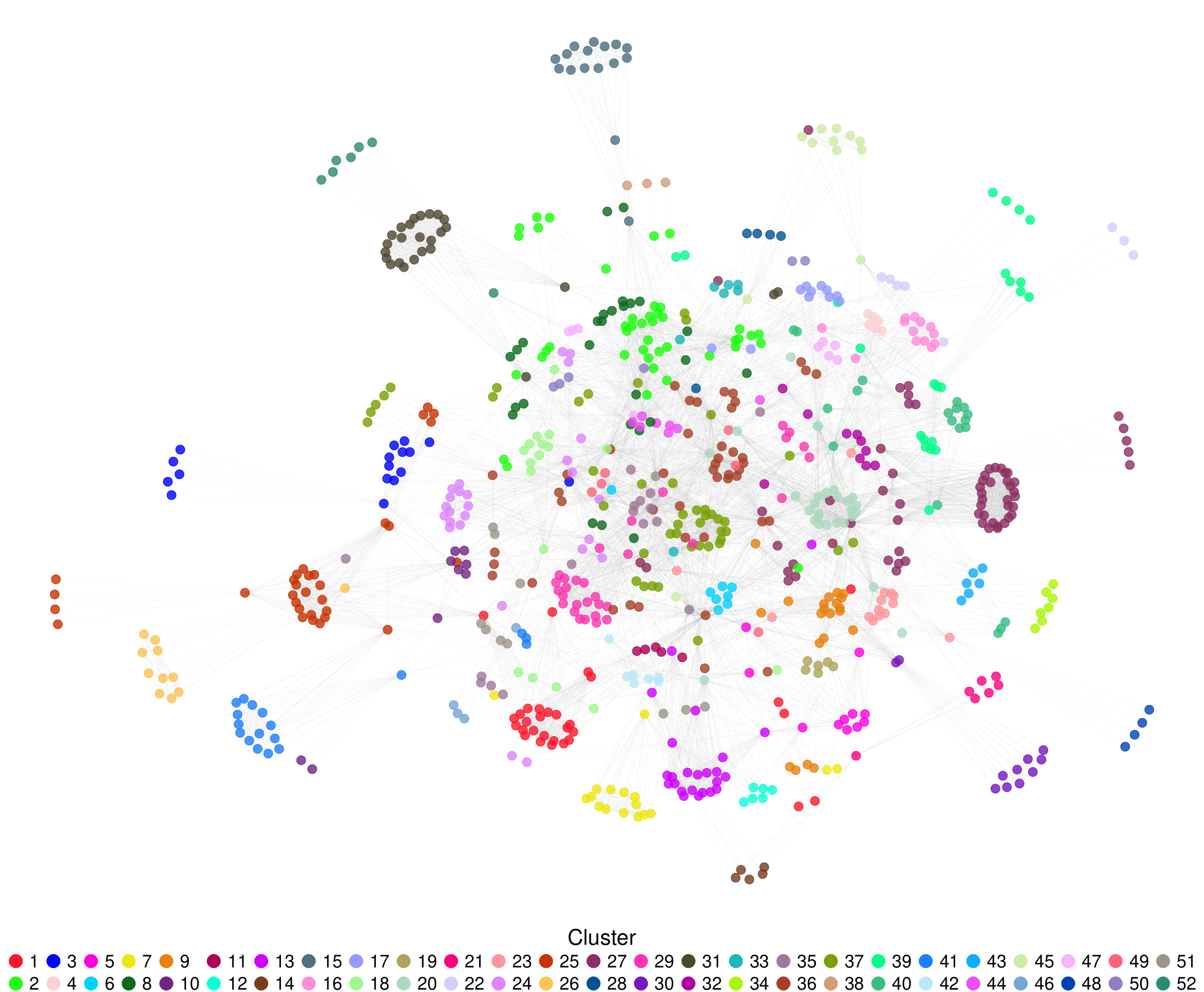}
        \caption{}
        \label{fig:sub11}
    \end{subfigure}
    % Second subfigure
    \begin{subfigure}[t]{.5\textwidth}
        \centering
        \includegraphics[width=\textwidth]{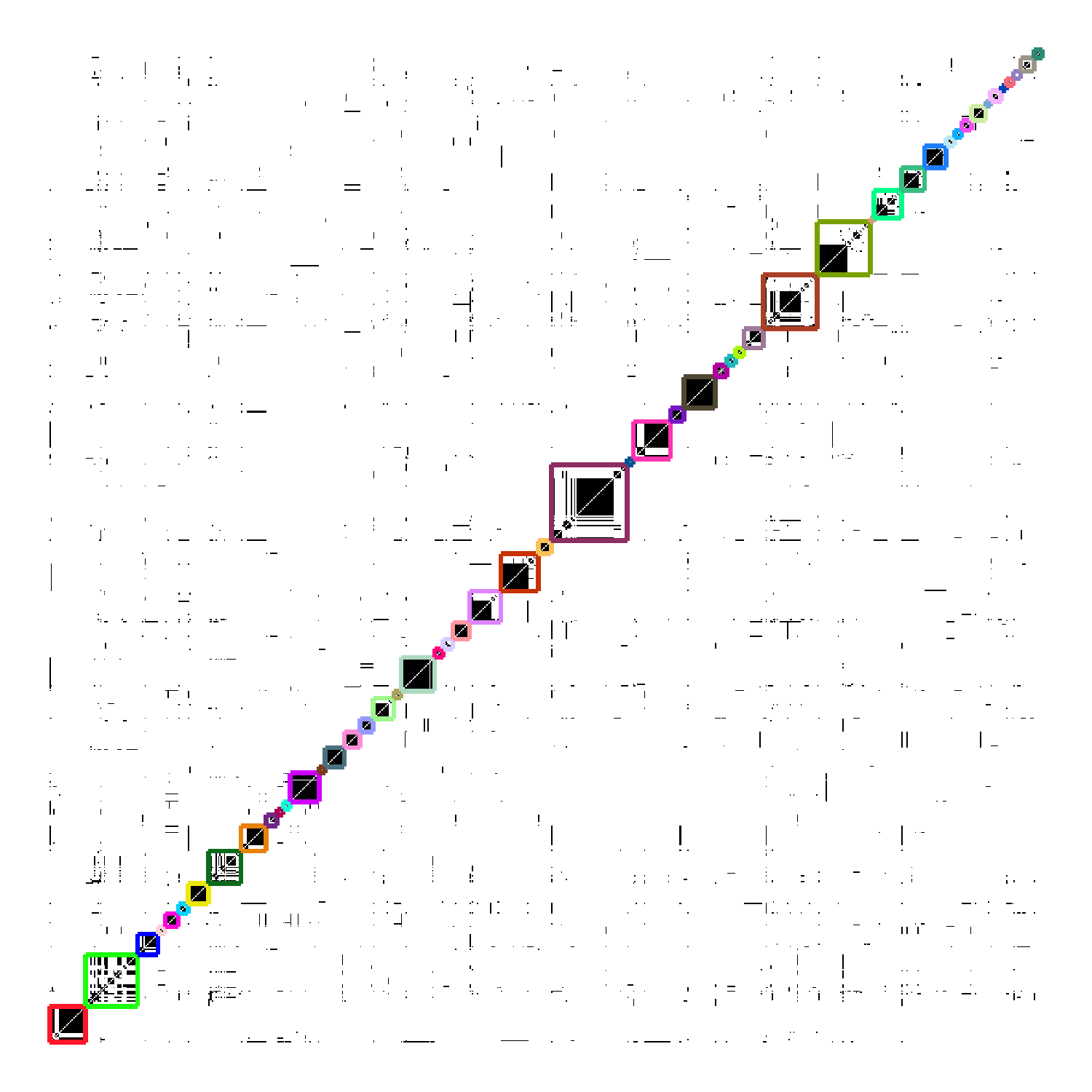}
        \caption{}
        \label{fig:sub12}
    \end{subfigure}

    \caption{Horizon Hydrogen network partition in communities using AttDeCoDe with kNN density on the project keyword embeddings. (a) Horizon network (b) Block-diagonal adjacency matrix representing intra and inter-community links.}
    \label{fig:fig8}
\end{figure}

When AttDeCoDe is applied using the GMM-based density, the method yields 11 clusters, dominated by one large component of 767 organisations and a set of very small, highly isolated groups--often behaving as near-cliques. Among these, only two of the ten small clusters contain connected organisations participating in at least two distinct projects. These clusters exhibit thematic coherence (e.g., hydrogen materials, electrolysis, mobility applications) and consistent institutional and geographic profiles, suggesting that GMM detects only the most concentrated local maxima in the attribute space. However, its inability to distinguish the broader low-density regions of the space limits its interpretability. In contrast, AttDeCoDe using the kNN-based attribute density (Figure \ref{fig:fig7}) provides a more balanced and informative representation, partitioning the organisations into 52 dense communities (Figure \ref{fig:fig8}).

\subsection{Characterisation of the detected clusters}

The clusters, obtained using AttDeCoDe with kNN, display a highly uneven size distribution, ranging from small groups of 3–5 organisations to large clusters comprising more than 60 members. The majority of clusters are moderately sized (10–30 organisations), reflecting the coexistence of compact, cohesive subgroups and broader, cross-sectoral alliances typical of Horizon consortia. This heterogeneity in size suggests that community formation in the Horizon network operates at multiple scales, with both tightly focused collaborations and large, integrative research domains coexisting within the same system. 

The correspondence between cluster size and project participation further illustrates this diversity. While nine clusters include organisations participating in a single project only, a small number of large clusters aggregate a substantial number of projects--up to more than 50 in the largest case (cluster 36 in Figure \ref{fig:fig8}). This pattern indicates the presence of broad research domains that integrate multiple collaborative initiatives and act as structural backbones of the network.

Tables \ref{tab:cluster_country} and \ref{tab:cluster_topic} provide a detailed quantitative characterisation of the clusters ranked among the top five by size (with ties) in terms of geographical composition, activity type, and dominant research topics. 
The composition of the clusters is consistent with the overall structure of the dataset and with the design principles of Horizon consortia. Geographically, these clusters are generally transnational, involving organisations from several highly represented countries--most notably France, Italy, Germany, and Spain. These results are consistent with \cite{Balland2019Network20032017}, who show that EU Framework Programmes foster highly integrated, transnational research networks, with certain organisations acting as key connectors between countries, reflecting patterns of collaboration similar to those observed in our clusters. However, the degree of concentration varies in other identified clusters. For example, cluster 25 is largely national, being dominated by Greek organisations (20 out of 31), whereas clusters 27, 36, and 37 display a more balanced international composition. This pattern indicates that some research areas remain embedded in national ecosystems, while others operate at a European scale. 
In terms of activity type, large clusters are predominantly composed of private research companies, reflecting the applied orientation of Horizon projects, but they systematically include research organisations and higher-education institutions. This cross-sectoral mix aligns with Horizon programme rules, which promote partnerships across countries and activity types, particularly between academia and industry. The findings are in line with previous studies on EU-funded collaboration networks, which document spatial heterogeneity in community formation \citep{barber2013community, Morea2024MappingAnalysis}.

The dominant topics reported in Table~\ref{tab:cluster_country} further characterise the largest clusters. For each cluster, the table reports the most frequent project keywords (top five by rank, allowing for ties), with those associated with the cluster leader highlighted in bold. Leaders correspond to the cluster cores identified by AttDeCoDe and represent organisations located in high-density regions of the attribute space, around which the remaining cluster members are organised.

Across all large clusters, fuel cells and electrolysis emerge as recurrent transversal themes, reflecting the overarching focus of the funded research and the strategic priorities of European hydrogen policy. At the same time, each cluster is structured around a distinct combination of secondary topics, revealing meaningful thematic differentiation.

Cluster~36 combines \textit{electrolysis} with \textit{productivity}, \textit{geometry}, and mining-related topics, suggesting a focus on large-scale industrial processes and optimisation. Cluster~37 is centred on \textit{metallurgy} and \textit{thermodynamic engineering}, with a clear emphasis on applied engineering and materials science. Cluster~2 exhibits stronger links to \textit{catalysis}, organic compounds, and chemical processes, while cluster~29 integrates sustainability-oriented themes such as \textit{ecosystems} and \textit{sensors} with energy-related topics. Finally, cluster~25 is characterised by \textit{public transport} and \textit{digital electronics}, pointing to application domains related to mobility and infrastructure rather than to core hydrogen technologies alone.

This pattern--shared core themes coupled with differentiated thematic specialisation--is consistent with previous studies of EU-funded collaboration networks, which document the emergence of field-specific communities organised around common strategic priorities but diversified along technological and application-oriented dimensions \citep{barber2013community, Balland2019SmartDiversification}.

Beyond the large clusters, only 13 clusters contain topics that are unique to them. Among these, two clusters--37 and 52--are particularly distinctive, as in both cases organisations participate in projects for which more than half of the associated keywords are unique to that cluster. Cluster 37 exhibits a strong technological and industrial orientation, combining \textit{metallurgy}, \textit{thermodynamic engineering}, \textit{fuel cells}, and advanced manufacturing. This focus is directly reflected in the leader’s keyword profile, reinforcing the interpretation of this cluster as a cohesive engineering-oriented community.
In contrast, cluster 52 is markedly research-driven, focusing on \textit{tissue engineering}, \textit{stem cells}, \textit{piezoelectrics}, and chemical engineering. Its members are primarily academic and research institutions, mostly located in France, Belgium, and other associated countries. 
The leader of this cluster exhibits a broad yet coherent thematic profile--spanning \textit{chemical engineering}, \textit{coating and films}, \textit{electrolysis}, \textit{photocatalysis}, \textit{piezoelectrics}, \textit{solar energy}, \textit{stem cells}, and \textit{tissue engineering}--which captures the interdisciplinary nature of the community at the intersection of life sciences, materials science, and energy-related technologies.
Importantly, such a niche structure is not apparent from structural connectivity alone and emerges clearly only when attribute-based density is taken into account.

Overall, these results show that the proposed approach is able to recover communities that are not only structurally cohesive but also thematically interpretable. The identified leaders provide a clear thematic anchor for each community, facilitating interpretation of the underlying research focus. Large clusters correspond to broad, multidisciplinary research domains that integrate multiple projects, countries, and sectors, whereas smaller clusters capture more specialised thematic niches. The joint analysis of cluster size, organisational composition, thematic focus, and leader profiles thus offers a comprehensive view of how collaboration patterns in the Horizon network emerge from the interplay between structural connectivity and attribute-driven similarity.

Thanks to the proposed approach, it emerges that EU-funded hydrogen research collaborations form a multi-scale, cross-sectoral, and largely transnational landscape. Core themes such as \textit{fuel cells} and \textit{electrolysis} recur across clusters, while secondary topics differ between communities, reflecting both thematic specialisation and the strategic priorities of Horizon consortia. Within each cluster, identified leader organisations serve as thematic anchors around which other members are structured, ensuring that the network combines cohesion with specialised focus and captures the coexistence of European-scale integration and local research ecosystems.

\begin{table}[bt!]
\centering
\caption{Composition of the five largest clusters (including ties) by country and activity type}
\label{tab:cluster_country}
\resizebox{\textwidth}{!}{
\begin{tabular}{clll}
\toprule
Cluster & Cluster size & Main countries (N) & Activity types (N) \\
\midrule
27 & 63 &
FI (17), RoE (16), DK (8), DE (6) &
PRC (33), HES (10), RoE (10), REC (8) \\

36 & 45 &
DE (8), FR (8), BE (5), ES (5), IT (5) &
PRC (32), HES (8), REC (4) \\

37 & 44 &
ES (12), DE (8), FR (5), IT (5), NL (4) &
PRC (30), REC (7), HES (6)  \\

2 & 43 &
ES (9), RoE (8), DE (7), IT (6), NL (4) &
PRC (29), HES (10), REC (4)  \\

29 & 31 &
OTH (15), IT (13), BE (2) &
PRC (22), REC (5), HES (3)  \\

25 & 31 &
EL (20), RoE (3), NL (2), UK (2) &
PRC (19), REC (5), HES (3), PUB (2)  \\
\bottomrule
\end{tabular}
}
\end{table}

\begin{table}[bt!]
\centering
\caption{Composition of the five largest clusters (including ties) by dominant research topics. Project keywords associated with cluster leaders in bold.}
\label{tab:cluster_topic}
\resizebox{\textwidth}{!}{
\begin{tabular}{cl}
\toprule
Cluster & Top 5 keywords (N) \\
\midrule
27 & 
\textbf{natural gas} (43); electrolysis (25); fuel cells (17); catalysis (8); machine learning (8); synthetic fuels (8) \\

36 & 
\textbf{electrolysis} (33); \textbf{geometry} (21); \textbf{productivity} (21); catalysis (12); fuel cells (12); mining and mineral processing (12) \\

37 & 
\textbf{metallurgy} (25); \textbf{thermodynamic engineering} (25); electrolysis (17); fuel cells (14); coating and films (8) \\

2 & 
\textbf{catalysis} (27); \textbf{aliphatic compounds} (20); \textbf{lipids} (20); electrolysis (11); coating and films (10) \\

29 & 
\textbf{ecosystems} (25); \textbf{fuel cells} (25); \textbf{sustainable economy} (25); natural gas (8); sensors (8) \\

25 & 
\textbf{public transport} (23); \textbf{digital electronics} (23); sustainable economy (7); electrolysis (7); coating and films (6) \\
\bottomrule
\end{tabular}
}
\end{table}

\section{Conclusion}\label{sec:conclusion}

This paper has proposed an attribute-driven extension of the DeCoDe framework of \citet{Menardi2022Density-basedNetworks}, designed for settings in which group structure is shaped jointly by network topology and node attributes. The central methodological contribution is to shift the identification of leaders from purely structural prominence to high density in an attribute space, thereby allowing communities to be organised around nodes that are representative in terms of characteristics.
%The approach introduces an attribute-based notion of node local density and integrates it into DeCoDe. 
A complementary simulation design extends the DC-SBM \citep{Karrer2011StochasticNetworks} by combining block structure with attribute-driven degree heterogeneity, providing a principled environment to evaluate our novel approach against established alternatives. Across a wide range of sparse network configurations, incorporating attribute-driven density in the leader-identification step improves the community recovery relative to both purely structural and attribute-assisted competitors.

The empirical application further demonstrates the practical value of the method. In the Horizon collaboration network, AttDeCoDe uncovers thematically coherent communities and succeeds in distinguishing small, specialised thematic clusters from broader cross-sectoral groupings--patterns that are not easily identifiable when relying on structural centrality alone. The proposed method, it also allows to discover (thematic) leader organisations that, guiding the observed community structure, are crucial for the overall network cohesiveness. Furthermore, the identification of such actors may be crucial for funding distribution policies \citep{Morea2024MappingAnalysis}. %Indeed, DeCoDe based on degree collapses the network into a single dominant cluster, whereas DeCoDe based on local density fragments it into near-cliques. Attribute-driven density offers a clearer and more interpretable picture of how research themes, organisational roles, and collaboration patterns interact.

Although AttDeCoDe computes density exclusively from node attributes, structural information can still influence which nodes emerge as modal actors. This is particularly evident in our bipartite setting, where event-level information projected onto nodes implicitly encodes participation structure: organisations involved in many events, or in events with similar thematic profiles, accumulate attributes that place them in denser regions of the feature space. As a result, the attribute representation captures aspects of structural prominence.

More generally, this mechanism arises when attributes carry relational content, such as interaction weights or other attribute-based measures of collaboration intensity. For example, quantities such as the total monetary value of the projects in which an organisation participates \citep{Morea2024MappingAnalysis} could likewise be incorporated as node-level density indicators by computing node strengths—defined as the sum of monetary values of the project collaborations incident to each organisation—thereby providing an alternative representation of leadership in collaborative settings. Likewise, node embeddings obtained through representation learning techniques--whether based on random walks, spectral methods, or graph neural networks--can be constructed by fusing structural and attributive information \citep{cai2018, cui2018}. 
These examples illustrate natural extensions of the AttDeCoDe framework. More generally, they highlight a key advantage of AttDeCoDe: although density is formally defined in the attribute space, the method remains sensitive to structural variation whenever relational information is directly or indirectly embedded in the node attributes.

%Overall, the results indicate that AttDeCoDe offers a valuable complement to probabilistic and spectral methods for attributed networks \citep[e.g.][]{Binkiewicz2017,Hu2024Network-adjustedDetection}, particularly in applications where heterogeneous degrees, overlapping dense cores, and thematic differentiation coexist. Beyond methodological novelty, the approach provides substantive insight into the mechanisms that shape collaborative structures, making it particularly well-suited to empirical applications in the social sciences.

Beyond these examples, the proposed methodology opens several additional avenues for further research. First, our implementation assumes a static, unweighted network. Extending AttDeCoDe to weighted or temporal networks \citep{Balland2019Network20032017, Morea2024MappingAnalysis} would broaden its applicability, enabling a better understanding of how communities evolve as collaborations expand, consolidate, or reorganise over time.
While the DeCoDe framework can accommodate weighted networks, doing so is non-trivial: empirical weight distributions are typically highly skewed and reflect factors such as consortium size, visibility, or resource availability, rather than purely structural interaction.

Second, the method inherits sensitivity to the choice of the density estimator (e.g.\ the number of mixture components in GMMs or the choice of $k$ in kNN) and in the construction of high-dimensional attributes (such as sentence embeddings for textual data). A systematic study of tuning strategies, uncertainty quantification, and robustness diagnostics would therefore be valuable for users. 

Third, while our simulation design relies solely on GMM-simulated attributes, this choice reflects both the motivating application and the common use of GMMs to model textual embeddings \citep{clinchant2013}.
%availability of textual data from which numerical embeddings are extracted--data commonly modelled using GMMs \cite{clinchant2013}. 
Nonetheless, the framework can incorporate alternative mixture models for which well-established density estimation theory exists, such as t-mixture models \cite{McLachlan2005}, allowing the exploration of settings with heavier tails or more heterogeneous attribute distributions. 

Lastly, the presented empirical application focuses on a relatively narrow research domain relevant to innovation dynamics and relies exclusively on the CORDIS database. This limits the scope to EU-funded initiatives and may overlook collaborations supported by national programmes, regional authorities, or private investment. Integrating information from alternative funding sources would offer a broader view of the research ecosystem, though comparable datasets are not currently accessible.

Despite these caveats, the proposed framework shows that attributive information can be integrated into DeCoDe in a computationally tractable manner, detecting clusters that are both structurally cohesive and thematically interpretable. We anticipate that AttDeCoDe will be useful in a range of applications where community formation is driven by the joint influence of network structure and node characteristics, such as scientific collaboration, political alliances, and organisational ecosystems, and that it will serve as a starting point for further methodological developments at the interface of network statistics and multivariate density estimation.
Overall, AttDeCoDe offers a flexible and interpretable framework for community detection in attributed networks, well suited to large-scale collaboration systems shaped by the joint influence of leadership and homophily.

\clearpage

\section*{Acknowledgements}

The authors acknowledge financial support under the National Recovery and Resilience Plan (NRRP), with European Union resources, NextGeneration EU -- National Recovery and Resilience Plan, Mission 4 -- Component 1 -- Investment 4.1 -- Project Title Models and methods for the analysis of collaboration networks -- CUP J53D23011540006.\\

The doctoral scholarship is co-financed with European Union resources, NextGeneration EU -- National Recovery and Resilience Plan, Mission 4 -- Component 1 -- Investment 4.1 -- CUP J92B22000900007. \\

The authors thank Agenzia Lavoro \& SviluppoImpresa Friuli Venezia Giulia for the support.

\clearpage

\bibliographystyle{apalike}
\bibliography{references}

\clearpage

% Appendix ---------------------------------------------------------------------------
\appendix

\section{Details of data generation in simulation study}\label{sec:appendix:sim_details}

\setcounter{table}{0}
\renewcommand{\thetable}{A\arabic{table}}

We explore multiple network sizes, $n = \{50, 150, 300\}$, and numbers of communities, $K = \{5, 10 \}$. 
%\subsection{Community assignment generation}
Nodes are assigned to communities according to a probability distribution
$\boldsymbol{\tau} = (\tau_1, \ldots, \tau_k, \ldots, \tau_K)$. This probability distribution is defined in two alternative ways:

\begin{itemize}
    \item \textbf{Uniform}

    Each community is equally likely, hence
    $$\tau_k = \frac{1}{K}, \qquad k = 1,\ldots,K.$$

    \item \textbf{Non-uniform}

    The community proportions are drawn from a symmetric Dirichlet distribution:
    $$\boldsymbol{\tau} \sim \mathrm{Dirichlet}(\mathbf{1}_K).$$
\end{itemize}

Node labels are then drawn as
    $$q_i \mid \boldsymbol{\tau} \sim \mathrm{Categorical}(\boldsymbol{\tau}).$$

Degree heterogeneity is introduced through node-specific parameters
$\hat\gamma_i$, derived from the attribute densities produced by the GMM, with $\overline{\hat\gamma}$ being the mean attribute density.
Let $\delta_i$ denote the density of node $i$ derived from $\hat\gamma_i$.
We first apply an exponential transformation:

$$
\gamma^{*}_i = \exp\!\left(\frac{\hat\gamma_i}{{\overline{\hat\gamma}}}\right),
$$
which amplifies the differences between high- and low-density nodes.
We then derive node density $\delta_i$ rescaling the transformed $\gamma^{*}_i$ to the $[0,1]$ interval:

$$
\delta_i = 
\frac{\gamma^{*}_i - \min(\bm\gamma^{*})}{\max(\bm\gamma^{*}) - \min(\bm\gamma^{*})}.
$$

To prevent low-density nodes from becoming isolated, we impose a minimum within-community bound on degree parameters. To maintain controlled sparsity, within-community edge probabilities scale with $\log(n)/n$, ensuring that, as $n\to\infty$, the expected degree remains of order ${\cal O}(\log n)$ to preserve both community detectability and connectivity in sparse network models. We set
$$
\rho_{\min} = 5 \cdot \frac{\log n}{n},
$$
and then rescale
$$
\delta_i \leftarrow \delta_i(1 - \rho_{\min}) + \rho_{\min},
$$
which guarantees $\delta_i \in [\rho_{\min},1]$ and yields realistic leader–follower heterogeneity while preserving sparsity.

Network edges are generated via a DC--SBM.
Within community $q$, the probability that nodes $i$ and $j$ form a link is proportional to
$$
\Pr(i \leftrightarrow j \mid C(v_i)=C(v_j)=q) \propto \delta_i \delta_j.
$$

Nodes with larger $\delta$ therefore act as local leaders, attracting more intra-community ties.

In summary, this simulation design combines:
\begin{enumerate}
    \item \textbf{Attribute-driven heterogeneity:}
    nodes located in dense regions of attribute space become leaders with larger $\delta_i$.
    \item \textbf{Sparse community structure:}
    connectivity probabilities scale like $\log(n)/n$, producing realistic networks as size increases.
    \item \textbf{Flexible cluster imbalance:}
    community sizes can be uniform or drawn from a Dirichlet distribution.
\end{enumerate}

The resulting networks exhibit community structure driven jointly by discrete block membership and continuous attribute-based attraction, providing a realistic setting to evaluate whether the proposed AttDeCoDe algorithm can recover the true partition.

\section{Additional results of simulation study}\label{sec:appendix:sim_add_results}

\setcounter{figure}{0}
\renewcommand{\thefigure}{B\arabic{figure}}

\begin{figure}[t!]
    \centering
        \centering
        \includegraphics[width=\textwidth]{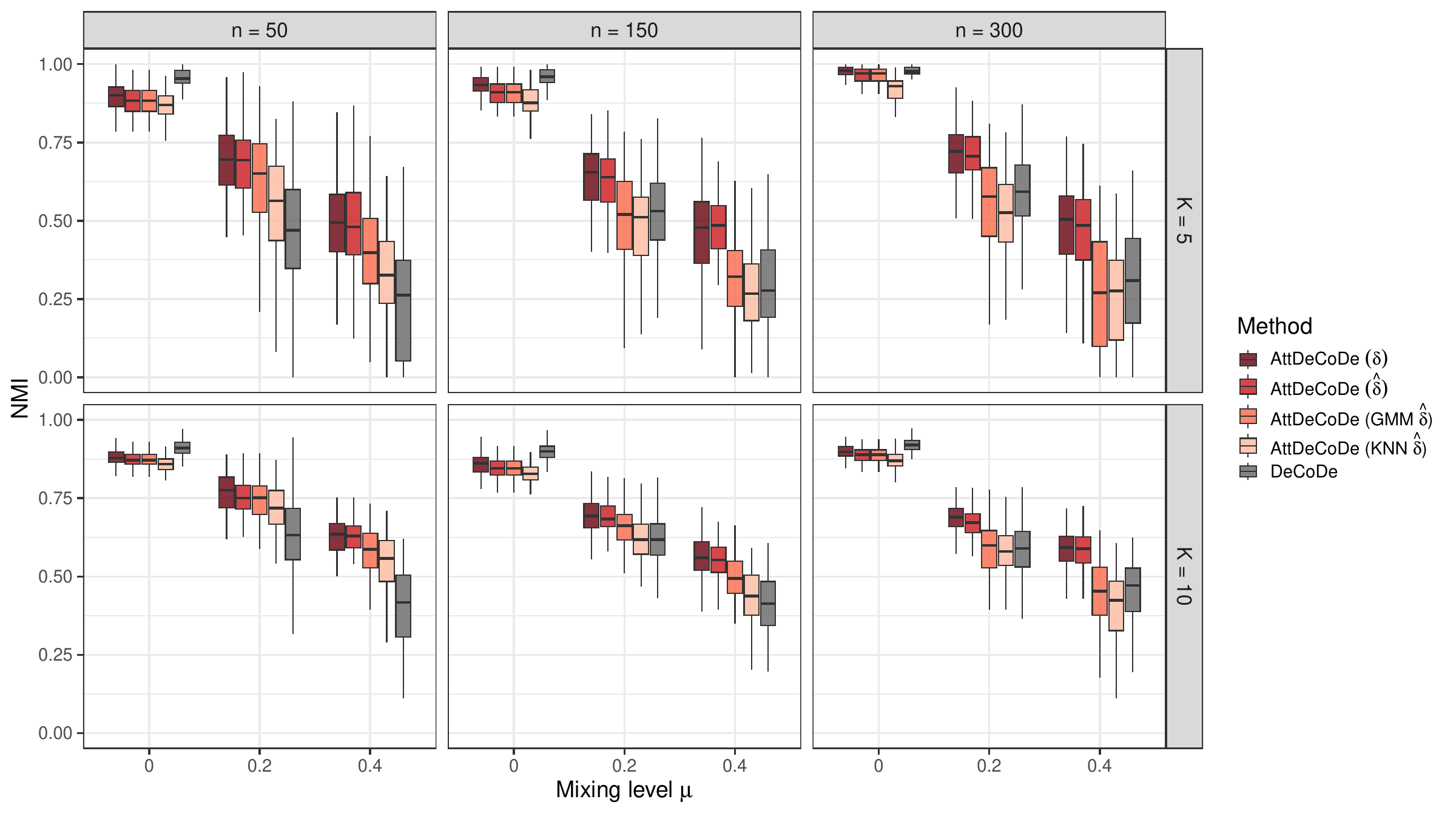}
        \caption{Normalised Mutual Information (NMI) distribution across different network sizes and numbers of communities ($K$) for \textbf{uniform} community sizes. Results are shown for AttDeCoDe (all density estimators) and DeCoDe.}

    \label{appx:fig1}
\end{figure}

Figure \ref{appx:fig1} reports additional results for the simulation study of Section~\ref{sec:simulation}. Across most configurations, the GMM and kNN density specifications improve cluster detectability compared to the original DeCoDe, consistent with the AttDeCoDe results discussed in the main text.

\begin{figure}[t!]
    \centering
        \centering
        \includegraphics[width=\textwidth]{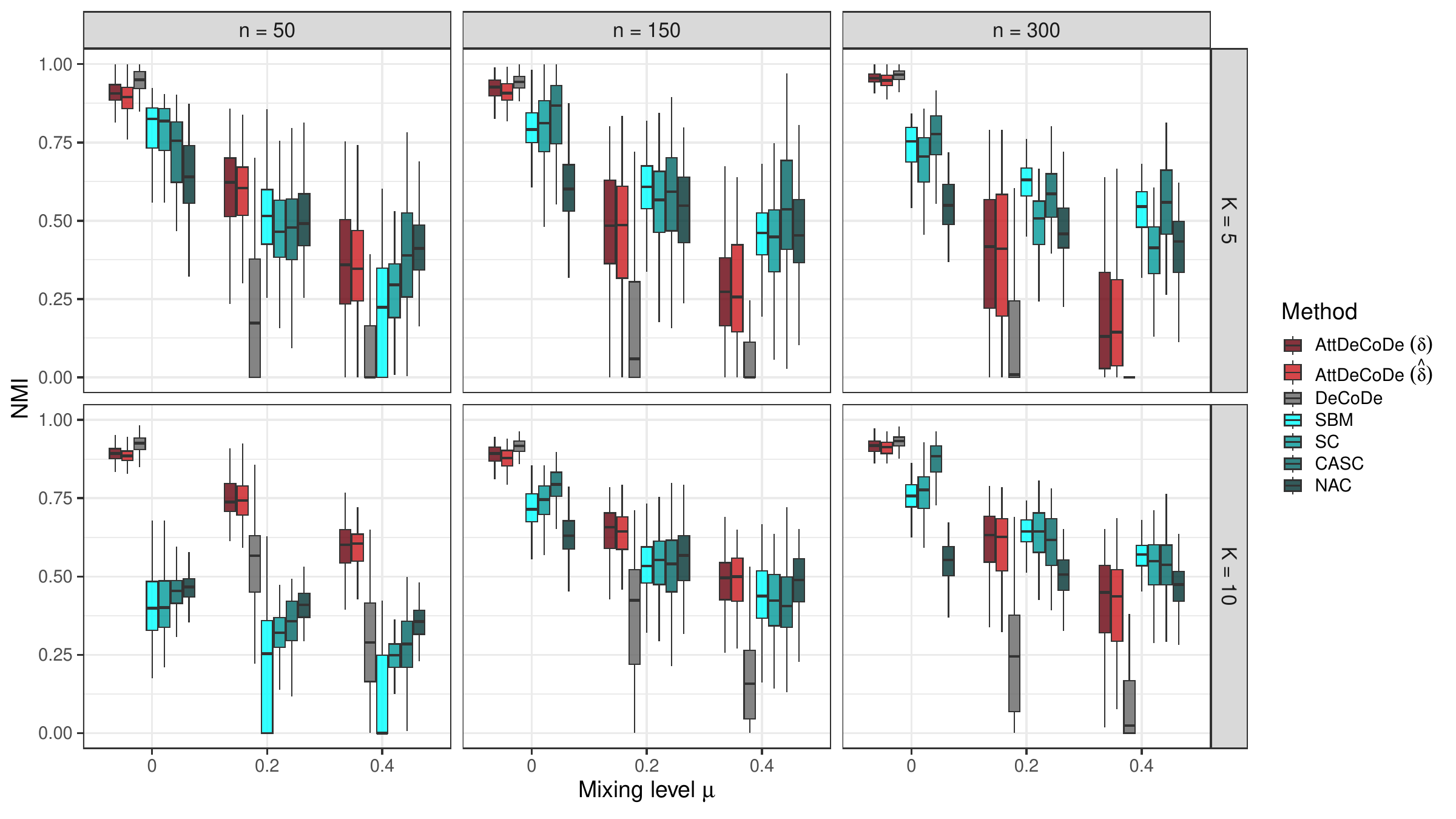}
        \caption{Normalised Mutual Information (NMI) distribution across different network sizes and numbers of communities ($K$) for \textbf{non-uniform} community sizes. Results are reported for all competing methods, including AttDeCoDe, DeCoDe, SBM, Spectral Clustering (SC), Covariate-Assisted SC (CASC), and SC on Network-Adjusted Covariates (NAC).}

    \label{appx:fig2}
\end{figure}

Figure \ref{appx:fig2} reports NMI values for DC-SBM networks with non-uniform community sizes. Overall, the patterns closely mirror those observed in the equal-size setting discussed in the main text, with the proposed AttDeCoDe variants consistently achieving high levels of agreement with the true partition. The primary difference lies in the increased variability of the results, which is particularly pronounced for the baseline methods.

\begin{figure}[t!]
    \centering
        \centering
        \includegraphics[width=\textwidth]{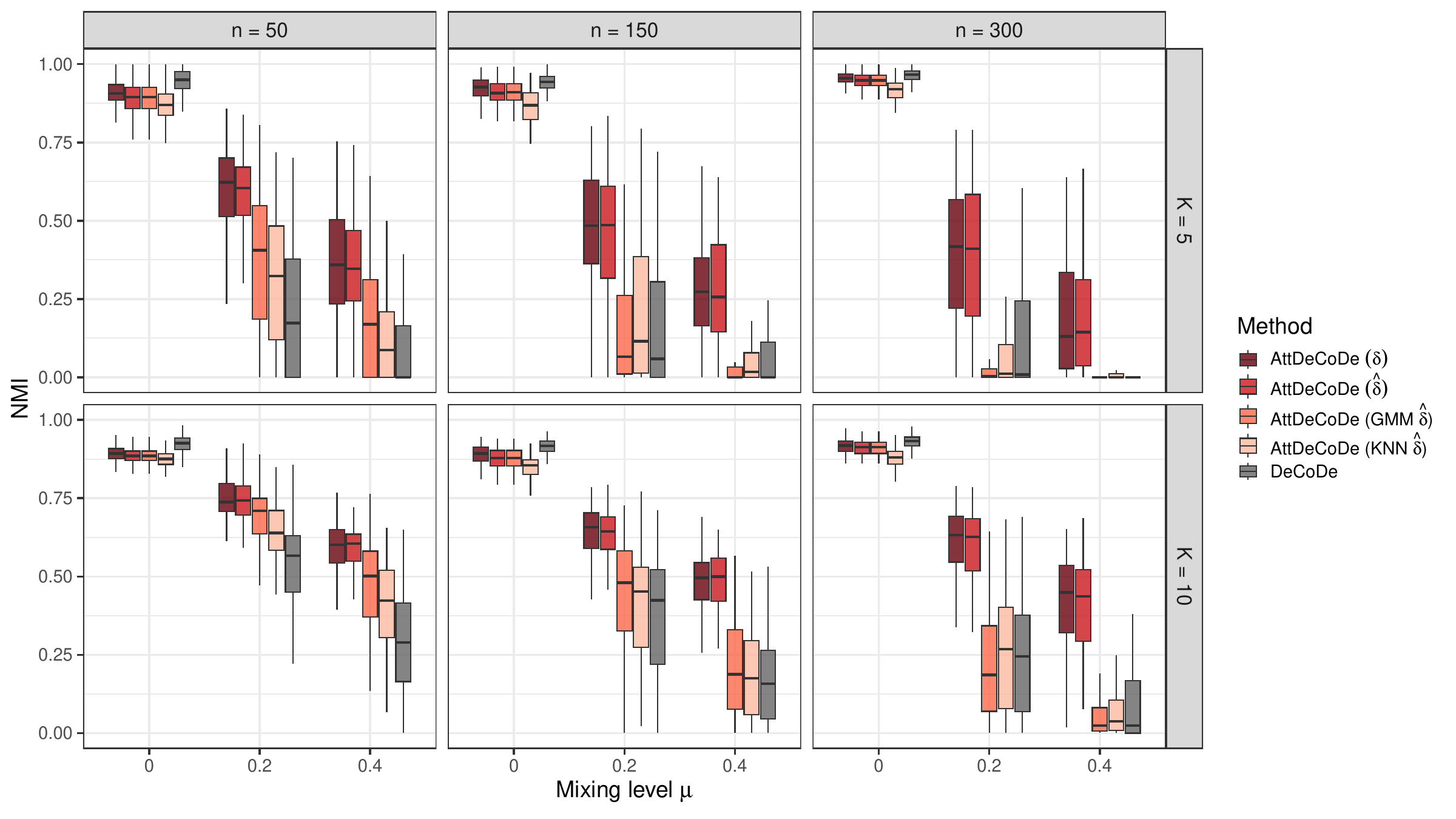}
        \caption{Normalised Mutual Information (NMI) distribution across different network sizes and numbers of communities ($K$) for \textbf{non-uniform} community sizes. Results are shown for AttDeCoDe (all density estimators) and DeCoDe.}

    \label{appx:fig3}
\end{figure}

Figure~\ref{appx:fig3} reports NMI values for networks with non-uniform community sizes, comparing AttDeCoDe under different density estimators with the original DeCoDe. Relative to the results shown in Figure~\ref{appx:fig1}, all methods exhibit greater variability and lower overall performance. This behaviour is consistent with the patterns observed in Figure~\ref{appx:fig2} and reflects the increased complexity of the setting, which allows for the presence of very small communities.

\end{document}